\documentstyle[twocolumn,aps,psfig]{revtex}
\begin{document}
\draft

\twocolumn[\hsize\textwidth\columnwidth\hsize\csname @twocolumnfalse\endcsname

\title{Hole Doping Evolution of the Quasiparticle Band in Models of
Strongly Correlated Electrons for the High-T$_c$ Cuprates}

%the 2D Hubbard
%Model with Next-Nearest-Neighbor Hopping}

\author{Daniel Duffy$^1$, Alexander Nazarenko$^2$, Stephan Haas$^3$, Adriana Moreo$^1$, 
Jose Riera$^4$ and Elbio Dagotto$^1$}
\address{$^1$ Department of Physics and National High Magnetic Field Lab, Florida State\\
University, Tallahassee, Florida 32306, USA}
\address{$^2$ Department of Physics, Boston College, Chestnut Hill, MA
02167, USA}
\address{$^3$ Theoretische Physik, Eidgen\"ossische Technische
Hochschule, 8093 Z\"urich, Switzerland}
\address{$^4$ Instituto de Fisica Rosario, Avenida 27 de Febrero 210 bis,
2000 Rosario,\\ Argentina}
\maketitle

\begin{abstract}

Quantum Monte Carlo (QMC) and Maximum Entropy (ME) techniques are used to study
the spectral function $A({\bf p},\omega)$ of the one band Hubbard model
in strong coupling
including a next-nearest-neighbor electronic
 hopping with amplitude $t'/t= -0.35$.
These values of parameters are chosen to improve the comparison of the
Hubbard model  with 
angle-resolved photoemission (ARPES) data for ${\rm Sr_2 Cu O_2 Cl_2}$.
A narrow quasiparticle (q.p.)
band is observed in the QMC analysis
at the temperature of the simulation $T=t/3$,
both at and away from half-filling. 
%At a small hole density the
%quasiparticle dispersion  resembles the result in the non-interacting limit
%$U/t=0$, although with reduced effective
%hopping amplitudes.
Such a narrow band produces a large accumulation of weight  in the
density of states at the top of the valence band. As the electronic
density $\langle n \rangle$ decreases further
away from half-filling, the chemical potential travels through
this  energy window
with a large number of states, and by $\langle n \rangle \sim 0.70$ it
has crossed it entirely. The region near momentum $(0,\pi)$ and $(\pi,0)$
in the spectral function
is more sensitive to doping than  momenta along the diagonal from
$(0,0)$ to $(\pi,\pi)$. The evolution with hole density of the quasiparticle
dispersion contains some of the features observed in
recent ARPES data in the underdoped regime. For sufficiently
large hole densities the ``flat'' bands at $(\pi,0)$
cross the Fermi energy, a prediction that could be tested with
ARPES techniques applied to overdoped cuprates.
The population of the q.p. band introduces a {\it hidden}
density in the system which produces interesting consequences when
the quasiparticles are assumed to 
interact through antiferromagnetic fluctuations
and studied with the BCS gap equation formalism. In particular,
a region of extended s-wave is found to compete with d-wave in the
overdoped regime, i.e. when the chemical potential has almost entirely
crossed the q.p. band as $\langle n \rangle$ is
reduced. The present study also shows that previous
``real-space'' pairing theories for the cuprates, such as the
Antiferromagnetic Van Hove scenario, originally
constructed based on information gathered at half-filling,
do not change  their predictions if
 hole dispersions resembling non-interacting electrons
with renormalized parameters are used.
%, as provided by 
%the QMC-ME results.

\end{abstract}


\vskip2pc]
\narrowtext

\section{Introduction}

Recent studies of 
${\rm Bi_2 Sr_2 Ca_{1-x} Dy_x Cu_2 O_{8+\delta}}$ 
using angle-resolved photoemission techniques have
provided the evolution of the quasiparticle band
as the hole density changes
in the underdoped regime of 
the high-$T_c$ cuprates\cite{marshall,ding,onellion1}.
These studies complement previous ARPES analysis of the
hole dispersion in the antiferromagnetic insulator 
${\rm Sr_2 Cu O_2 Cl_2}$\cite{wells,onellion3}. The
overall results emerging from these experiments can be
summarized as follows: (i) the bandwidth of the quasiparticle band
is a fraction of eV, i.e. narrower than generally predicted
by band structure calculations. This result suggests that
strong correlations are important in the generation of the
quasiparticle dispersion; (ii) the two dimensional (2D) $t-J$ model
explains accurately the bandwidth for a hole injected
in the insulator, as well as the details of the dispersion along
the main diagonal in momentum space from ${\bf p}=(0,0)$ to $(\pi,\pi)$
(in the 2D square lattice notation)\cite{wells}; (iii) however, the behavior
along the direction from $(0,\pi)$ to $(\pi,0)$
is not properly described by the $t-J$ model which
predicts a near degeneracy between momenta along this line\cite{review}, 
%In other words
%$(\pi/2,\pi/2)$ and $(0,\pi)-(\pi,0)$ are predicted to be very close in
%energy,\cite{review} 
in contradiction with experiments. The ARPES data for
the antiferromagnetic insulator show that the quasiparticle signal at $(\pi,0)$
is very weak, and about 0.3 eV deeper in energy than
$(\pi/2,\pi/2)$\cite{wells};
(iv) as the density of holes grows the region near $(\pi,0)$
moves towards the chemical potential
which is approximately
reached at the optimal concentration\cite{marshall}, while
the main diagonal
$(0,0)-(\pi,\pi)$ is less affected;
(v) at optimal doping remarkably ``flat'' bands at $(\pi,0)$ near the Fermi energy
have been observed\cite{flat-exper}. 
%These
% are remarkable features of the ARPES data

The results of Marshall et al.\cite{marshall} have been
interpreted as providing evidence for ``hole pockets'' in the
underdoped regime
caused by  short-distance antiferromagnetic correlations, although other
explanations such as preformed pairs are also possible\cite{prefor,campu}.
%
%are also an alternative explanation\cite{marshall}).
%
%as well as those constructed around
% the existence of preformed pairs have  been mentioned
%as a possible origin for some of 
%the above described interesting features observed with ARPES\cite{marshall}.
As explained before, the standard $t-J$ model is not enough to
provide all the details of
the ARPES results at half-filling, and thus presumably it
cannot explain the data away from half-filling either.
However, the $t-J$ model is just one possible Hamiltonian
to describe the behavior of holes in an antiferromagnetic 
background. While the model is attractive for its simplicity,
there is no symmetry or renormalizability argument
signaling it as unique for the description of the cuprates.
For this reason  Nazarenko et al.\cite{nazarenko} recently introduced
extra terms in the $t-J$ Hamiltonian to improve the agreement with
experiments. Adding an electronic hopping along the diagonals of the
elementary plaquettes (with amplitude $t'/t=-0.35$) the results
for the insulator were improved since the position of the
quasiparticle at  $(\pi,0)$ proved to be
very sensitive to the strength of extra hole hopping terms in the model.
Actually,
the q.p. peak at this momentum
moves towards larger binding energies as a negative $t'/t$ grows in 
amplitude\cite{nazarenko}.
This idea has been  used by several other groups which,
in addition to hoppings along the plaquette diagonals,
have included
hoppings at distance of two lattice spacings with amplitude
$t''$
to obtain an even better
agreement with experiments\cite{others}.

The purpose of this paper is to report on results for the density
dependence of the 
quasiparticle dispersion corresponding to 
the one band Hubbard model
with a next-nearest-neighbor hopping $t'$ working
in the strong Coulomb coupling regime. It is expected that the
$U-t-t'$ and $t-t'-J$ models  produce qualitatively similar
physics at large $U/t$. As numerical technique
we use the Quantum Monte Carlo method supplemented by Maximum
Entropy  analysis.
The success of the $t-t'-J$ model at half-filling\cite{nazarenko,others}
and the recent availability of ARPES data in the underdoped
regime\cite{marshall,ding,onellion1}
prompted us to carry out this study. 
Note that although the numerical methods  used here are powerful, their
accuracy is limited and, thus, our results are mostly qualitative
rather than quantitative. Nevertheless, from the analysis of the present QMC data
and also comparing our results against those produced previously for the 
$t-J$ and Hubbard models here we arrive to the conclusion that at least
some features of the 
experimental ARPES
evolution of the hole dispersion can be  explained
using one band electronic models. Experimental predictions are made
to test the calculations. The presence of a narrow quasiparticle
band in the spectrum, and its crossing by the chemical potential
as the electronic density is reduced, are key features for the
discussions below.

The present paper is not only devoted to the analysis of QMC results
and its comparison with ARPES data, but it also
addresses the influence of hole doping in recently proposed theories
for the cuprates\cite{afvh}.
In these theories the hole dispersion is calculated at half-filling
and assumed to change only slightly as the hole density grows.
Hole attraction is assumed to be dominated
 by the minimization of antiferromagnetic (AF)
broken bonds, 
which implies the presence of an effective nearest-neighbor (NN)
density-density attraction proportional to the exchange $J$.
Superconductivity in the ${ d_{x^2 - y^2}}$ channel is natural in
this scenario\cite{afvh} due to the strong AF 
correlations\cite{review,doug}.
The interaction of holes is better visualized in
$real$ $space$, i.e. with pairing occurring when dressed holes
share a  spin polaronic cloud, as in the spin-bag mechanism\cite{bob}.
This real-space picture (see also Ref.\cite{joynt}) 
holds even for a small AF
correlation length, $\xi_{AF}$, and in this scenario
there is no need to tune
parameters to work very close to an AF instability as in other approaches.

To obtain quantitative information from these intuitive ideas, 
holes moving with a dispersion calculated 
using one hole in an AF background, $\epsilon_{AF}({\bf p})$,
and interacting through the NN attractive potential
mentioned before, have been previously
analyzed\cite{afvh}.
Within a rigid band
filling of $\epsilon_{AF}({\bf p})$ and using a BCS formalism,
${ d_{x^2 - y^2}}$ superconductivity dominates with
$T_c \sim 100K$ caused by a large density of states
(DOS) that appears in the hole dispersion.
The idea has many similarities with
previous scenarios that used van Hove (vH) singularities 
in the band structure to increase
$T_c$\cite{vh}, although d-wave superconductivity
 is not natural in this context unless 
AF correlations are included.
However, the rigid band filling is an approximation whose accuracy
remains to be tested. In particular,
the following questions naturally arise:
(i) does the q.p. peak in the DOS found at $\langle n \rangle =1$
survive a finite hole density?;
(ii) to what extent do the changes in the q.p. dispersion with doping
affect previous calculations in this framework?;
(iii) are the ``shadow'' regions generated by
AF correlations\cite{bob,shadow,shadow2,onellion2} important for
real-space pairing approaches? With the help of the present QMC
results, as well as previous simulations,
in this paper all these issues are
discussed. The overall conclusion is that as long as the
hole density is such that the 
$\xi_{AF}$ is at least of a couple of lattice spacings, the 
predictions of previous scenarios\cite{afvh} and other similar
theories remain qualitatively the same, in spite of substantial
changes occurring in the hole dispersion with doping.

The organization of the paper is as follows: in Sec.II the model
and details of the numerical method are discussed. Sec.III contains
the QMC-ME results both at half-filling and with a finite hole
density. In Sec.IV the results are discussed and compared with
ARPES experiments. Implications for real-space theories of high-$T_c$
are extensively studied in Sec.V. 
Results of previous publications are also used to
construct a simple picture for the behavior of electrons in
the Hubbard model.  Sec.VI contains a summary of the paper and its
experimental predictions.

\section{Model and Numerical Technique}

The one band Hubbard Hamiltonian with next-nearest-neighbor hopping is
given by

$$
{H=} -t{\sum_{\langle {\bf{ij}}\rangle,\sigma} 
(c^{\dagger}_{{\bf{i}},\sigma} c_{{\bf{j}},\sigma}+h.c.)}
-t'{ \sum_{\langle\langle {\bf{ij}} \rangle\rangle,\sigma}
(c^{\dagger}_{{\bf{i}},\sigma} c_{{\bf{j}},\sigma}+h.c.)}
$$

$$
+U{ \sum_{{\bf{i}}}(n_{{\bf{i}} \uparrow}-1/2)( n_{{\bf{i}}
\downarrow}-1/2)+\mu\sum_{{\bf{i}},\sigma}n_{{\bf{i}}\sigma}},
\eqno(1)
$$

\noindent where ${ c^{\dagger}_{{\bf{i}},\sigma} }$ creates an electron
at site ${ {\bf i } }$ with spin projection $\sigma$, ${
n_{{\bf{i}}\sigma} }$ is the number operator, the sum ${ \langle
{\bf{ij}} \rangle }$ runs over pairs of nearest-neighbor lattice sites,
${\langle\langle {\bf ij} \rangle \rangle}$ runs over pairs of lattice
sites along the plaquette diagonals, $U$ is the on-site Coulombic
repulsion, ${t}$ the nearest-neighbor hopping amplitude, $t'$ is the
plaquette diagonal hopping amplitude, and $\mu$ the chemical potential.
Throughout this study we will set $t=1$, $t'/t=-0.35$, and use periodic
boundary conditions.

Using standard QMC methods\cite{blank}, we have obtained the imaginary
time Green's functions at finite temperature.  The method of Maximum
Entropy\cite{gubernatis}  
was used to analytically continue the imaginary time
Green's functions to obtain the spectral weight function $A({\bf
p},\omega )$.  Previous studies using this method have 
concentrated on, e.g.,
the one-dimensional \cite{preuss} and two-dimensional Hubbard model at
and away from half-filling
\cite{shadow2,whiteandmore,preuss2,moreo}.  However, with the next-nearest-neighbor hopping
term only a few results have been obtained with this technique
\cite{duffy} since the additional hopping amplitude exacerbates the
 sign problem (which exists even at half-filling).

In this paper, we present a systematic study of the evolution of
the spectral function in the $U-t-t'$ model at electronic densities
 ranging from
half-filling $(\langle n \rangle = 1.0)$ to quarter-filling $(\langle n
\rangle = 0.5)$ on a $6 \times 6$ lattice at an inverse temperature of $\beta t=3
(T=0.33t)$.  The procedure we used to obtain these results is slightly
different from those used before \cite{moreo,duffy}, and we
believe that it provides more details to the spectral function.  First,
approximately 100,000 QMC measurement sweeps were taken at each density
(for a fixed $T$, $t'/t$, and $U/t$) to obtain accurate
statistics for the imaginary time Green's functions.  Once the spectral
function was obtained from the standard ME procedure, we used this
spectral function as a seed for further ME analysis with a reduced
coefficient of the entropy $\alpha$  \cite{gubernatis,alphafoot}.
%Finally, to establish the peak positions, amplitudes, and widths, we
%fitted $A({\bf p},\omega )$ to a set of Gaussian functions while
%keeping the overall spectral weight constant.

In order to work in the realistic strong coupling regime, sacrifices in
the lattice size and temperature had to be made due to the sign
problem\cite{footnote}. The same set of measurements shown below in 
Sec.III could have been
performed on an $8\times 8$ lattice.  However, it would have required a higher
temperature $(\beta t=2)$ to obtain good statistics on the imaginary
time Green's functions thereby washing out parts of 
the quasiparticle features
that become prominent at lower temperatures.  Since at half-filling and
$\beta t = 2$, the $6\times 6$ and $8\times 8$ lattices give
qualitatively similar
results and we were able to reach a lower temperature using the smaller
lattice, then we decided to use the $6\times 6$ cluster throughout the paper.

\begin{figure}[t]
\centerline{\psfig{figure=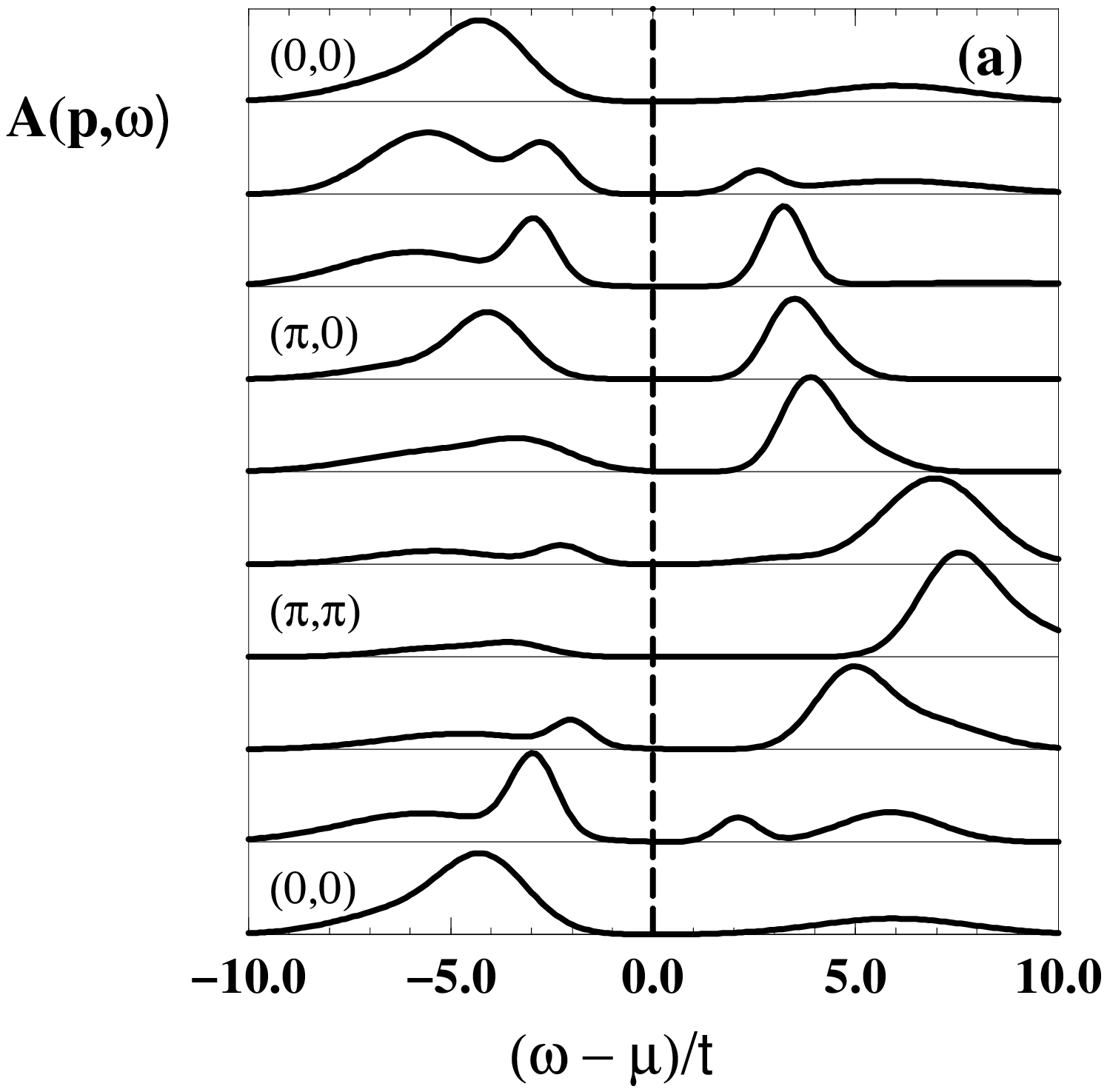,height=5.6cm,bbllx=132pt,bblly=155pt,bburx=570pt,bbury=578pt,angle=270}}
\centerline{\psfig{figure=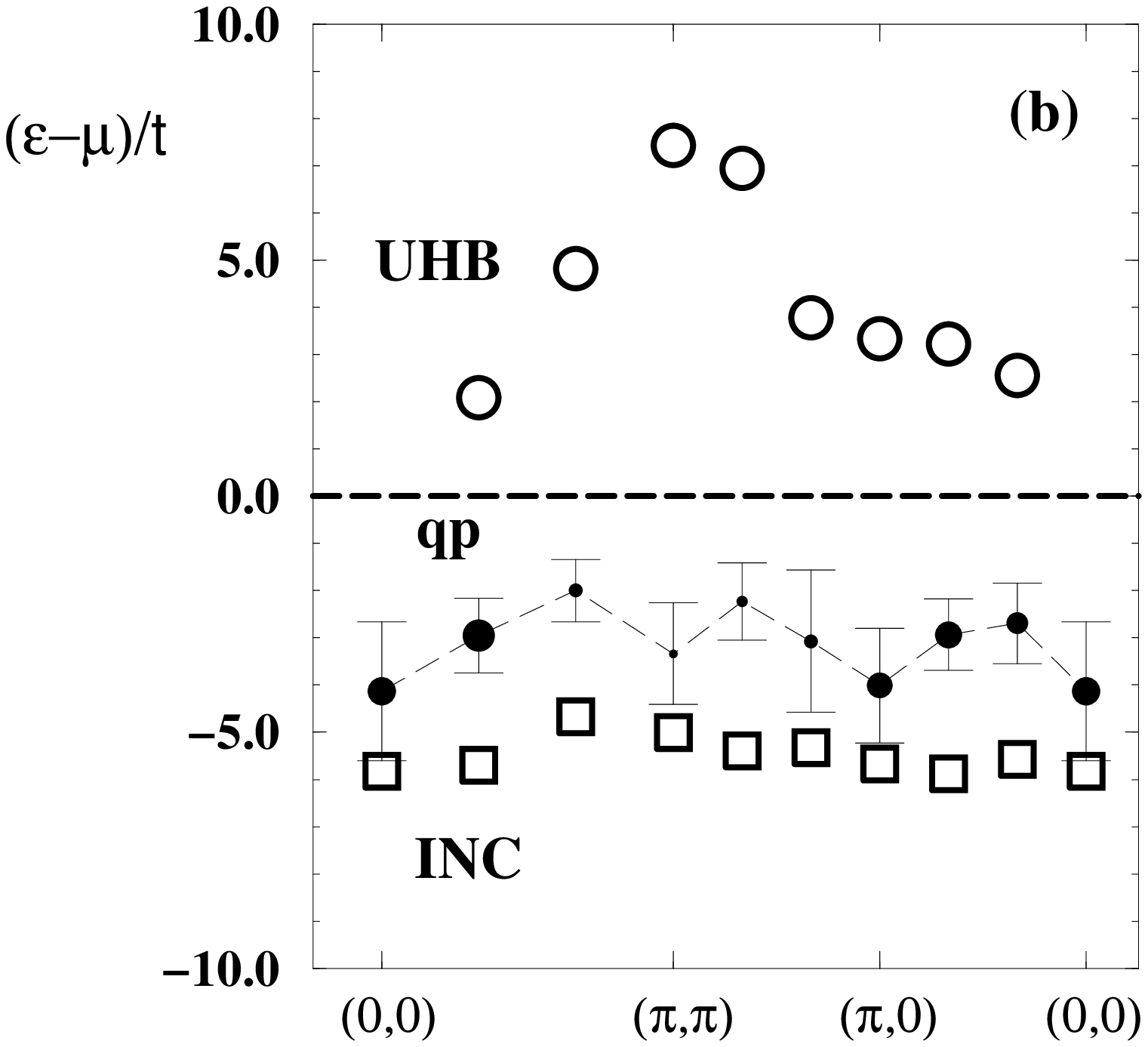,height=5.6cm,bbllx=132pt,bblly=155pt,bburx=570pt,bbury=578pt,angle=270}}
\centerline{\psfig{figure=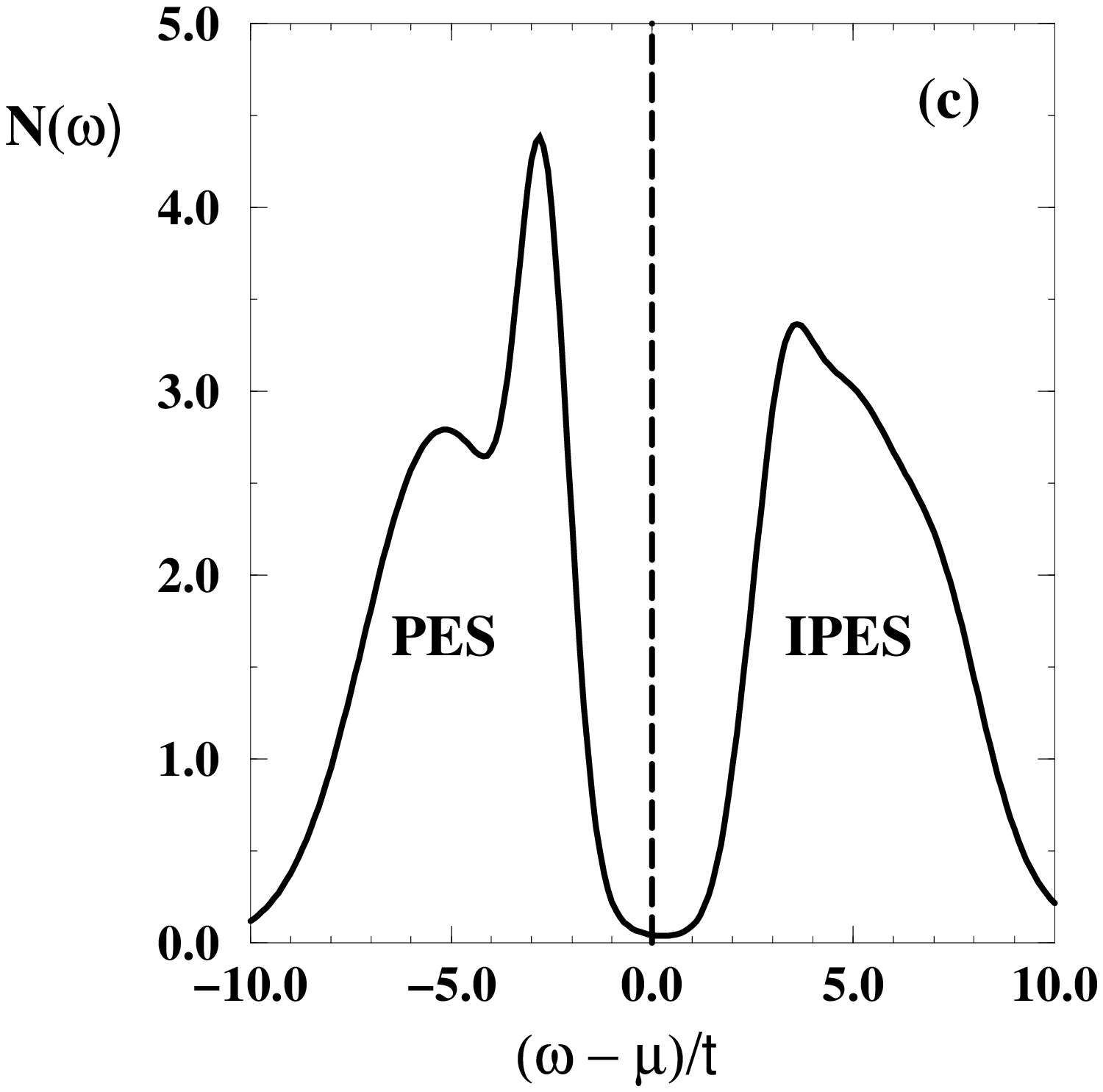,height=5.6cm,bbllx=132pt,bblly=155pt,bburx=570pt,bbury=578pt,angle=270}}
\caption{(a) Spectral functions $A({\bf p},\omega)$ of the $U-t-t'$ Hubbard
model at $U/t=10$, $t'/t=-0.35$, and $T=t/3$, using QMC-ME techniques on
a $6 \times 6$ cluster. The density is $\langle n \rangle = 1.0$. From
the bottom, the momenta are along the main diagonal from $(0,0)$ to
$(\pi,\pi)$, from there to $(\pi,0)$, and finally back to $(0,0)$; (b)
Energies of the dominant peaks in the spectral functions. The results in
the PES region are obtained from a two-gaussian analysis of the QMC-ME
results. The squares correspond to the incoherent part of the spectrum
(INC) and its error bars are not shown. The full circles form the
quasiparticle (qp) band, and their diameter is proportional to the
intensity. The error bars are given by the width at half the height of
the gaussian corresponding to the q.p.  peak at each momenta. The upper
Hubbard band is also shown (open circles) without error bars; (c)
density of states $N(\omega)$ obtained by summing the $A({\bf
p},\omega)$'s. The inset shows the top of the q.p. peak in more detail.}
\label{fig1}
\end{figure}

\section{QMC Results}

\subsection{Half-filling}

In Fig.1 the QMC results obtained at half-filling  for $U=10t$, $T=t/3$,
and $t'/t= -0.35$ are shown. Fig.1a
contains $A({\bf p},\omega)$ after the ME analysis of the QMC data.
The asymmetry of most of the dominant peaks suggests that they are
a combination of at least two features located at different energies.
This is reasonable since  previous calculations for the $U-t$
model (see e.g. Ref.\cite{moreo})  have established that below the chemical
potential $A({\bf p},\omega)$ is made out of a quasiparticle (q.p.) peak near $\mu$, carrying
a small fraction of the total weight, and a broad incoherent (INC)
 feature at larger binding energies containing the rest of the weight.
$A({\bf p},\omega)$ have been fitted by two gaussians with positions, widths
and weights adjusted to match the ME result, keeping the overall
weight constant. The same study
was performed above the
chemical potential in the inverse photoemission (IPES) 
regime but such analysis did not provide interesting
information since the weight of the peak the closest to $\mu$ is very
small, specially away from half-filling. Then,
within the accuracy of the present QMC-ME study,
the upper Hubbard band (UHB) is 
described as just containing a broad featureless peak for each momentum.
The two-peak decomposition analysis reported here will, thus, 
be limited
to the PES part of the spectrum. However, note that at least at
half-filling it is likely that the IPES portion of the spectrum contains
a (low intensity) q.p.-like
feature.
In particular as $t'/t \rightarrow 0$ particle-hole symmetry must be
recovered, and here the peaks appearing in the PES region must also exist
in the IPES regime. Nevertheless, in Sec. III.B below
 we will show that at finite
hole densities the q.p. band at the bottom of the UHB carries such a
small weight that it can be neglected.

Following this fitting procedure, 
in Fig.1b the energy position of the peaks 
as a function of momentum is shown. As anticipated, below $\mu$
a feature at the top of the valence band appears. It is natural to
associate this peak with a ``quasiparticle band''. The bandwidth 
is roughly $2t$, i.e. much smaller than for free electrons on a lattice.
However, it is not as small as predicted by $t-J$ model
calculations\cite{review},
which may be due to the influence of $t'$ or a finite $U/t$.
The top of the
q.p. band is located along the main diagonal from $(0,0)$ to
$(\pi,\pi)$, in agreement with previous literature\cite{review}.
The influence
of the nonzero next-nearest-neighbor (NNN) hopping $t'$ appears in the energy
position of the q.p. band at $(\pi,0)$ which is
deeper in energy than $(2\pi/3,2\pi/3)$ and
$(\pi/3,\pi/3)$ (while for $t'/t=0.0$, $(\pi,0)$ is
located very close to the top of the band in contradiction
with ARPES experiments for cuprate insulators\cite{wells}).
It is likely that on larger clusters
 $(\pi/2,\pi/2)$ 
would correspond to the actual top of the q.p. band, as
predicted before\cite{review}.
The overall shape of the q.p. band is similar to that calculated
in the $t-J$ model using the 
self-consistent Born approximation\cite{nazarenko}. 

\begin{figure}[t]
\centerline{\psfig{figure=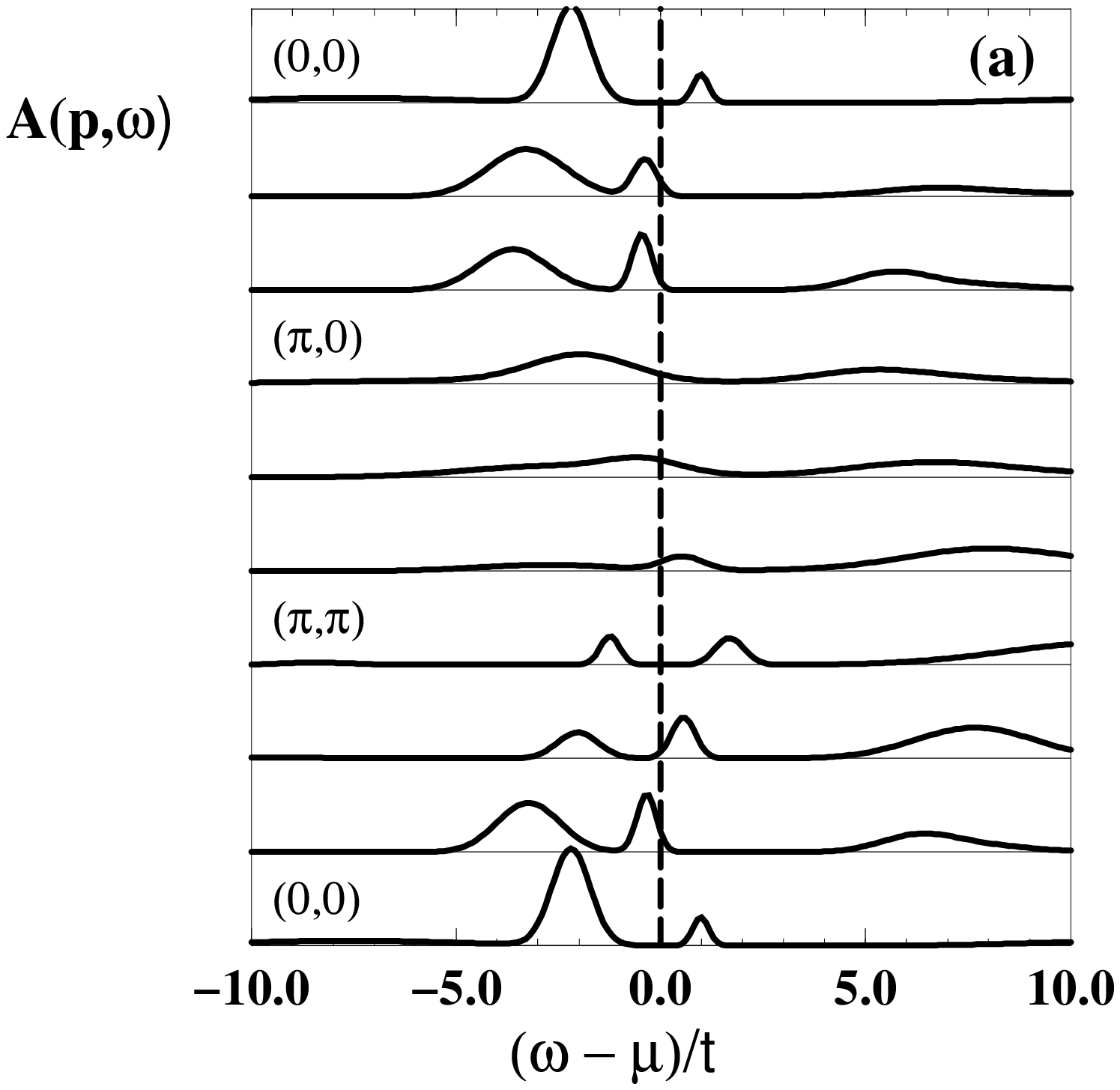,height=5.6cm,bbllx=132pt,bblly=155pt,bburx=570pt,bbury=578pt,angle=270}}
\centerline{\psfig{figure=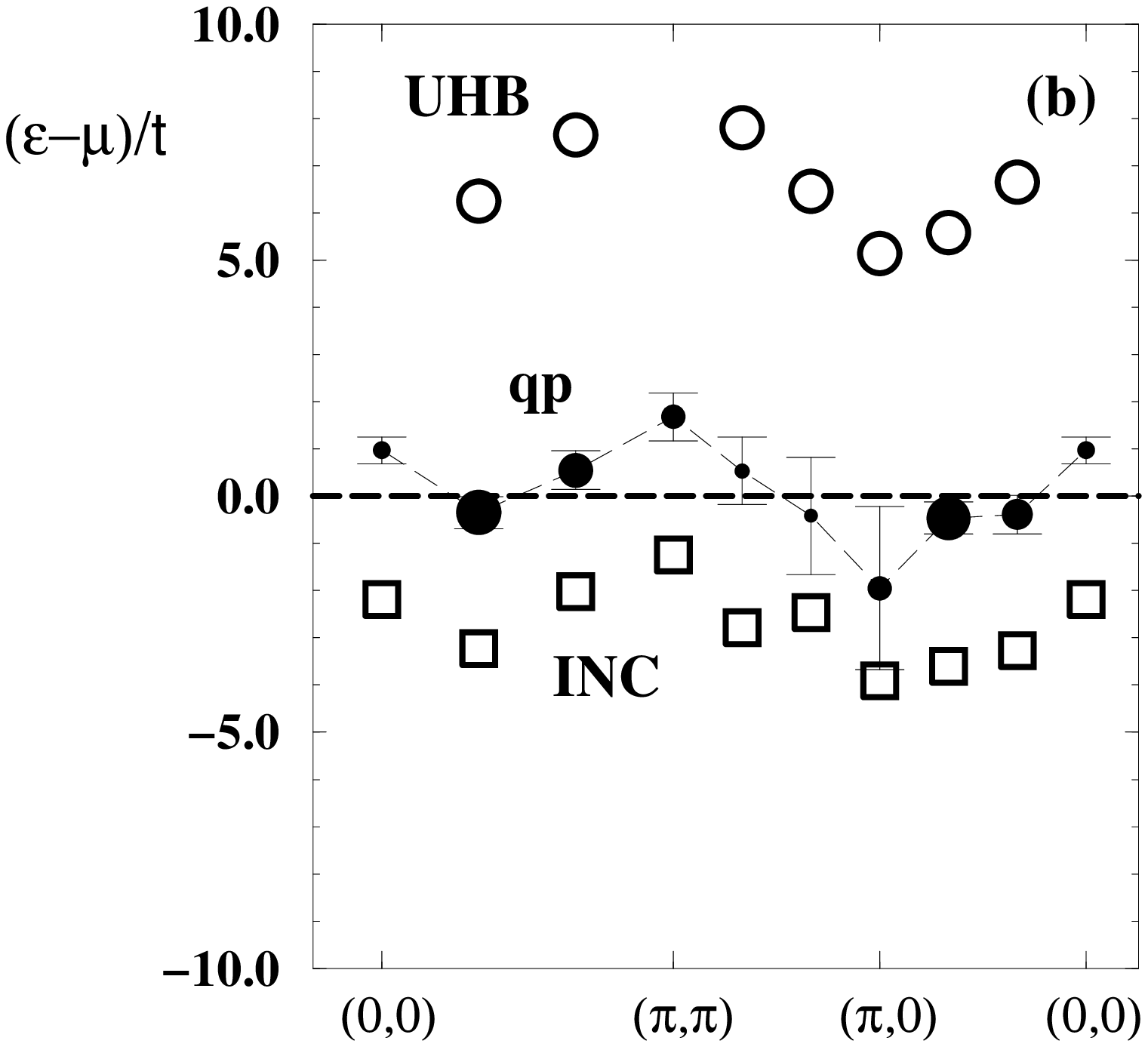,height=5.6cm,bbllx=132pt,bblly=155pt,bburx=570pt,bbury=578pt,angle=270}}
\centerline{\psfig{figure=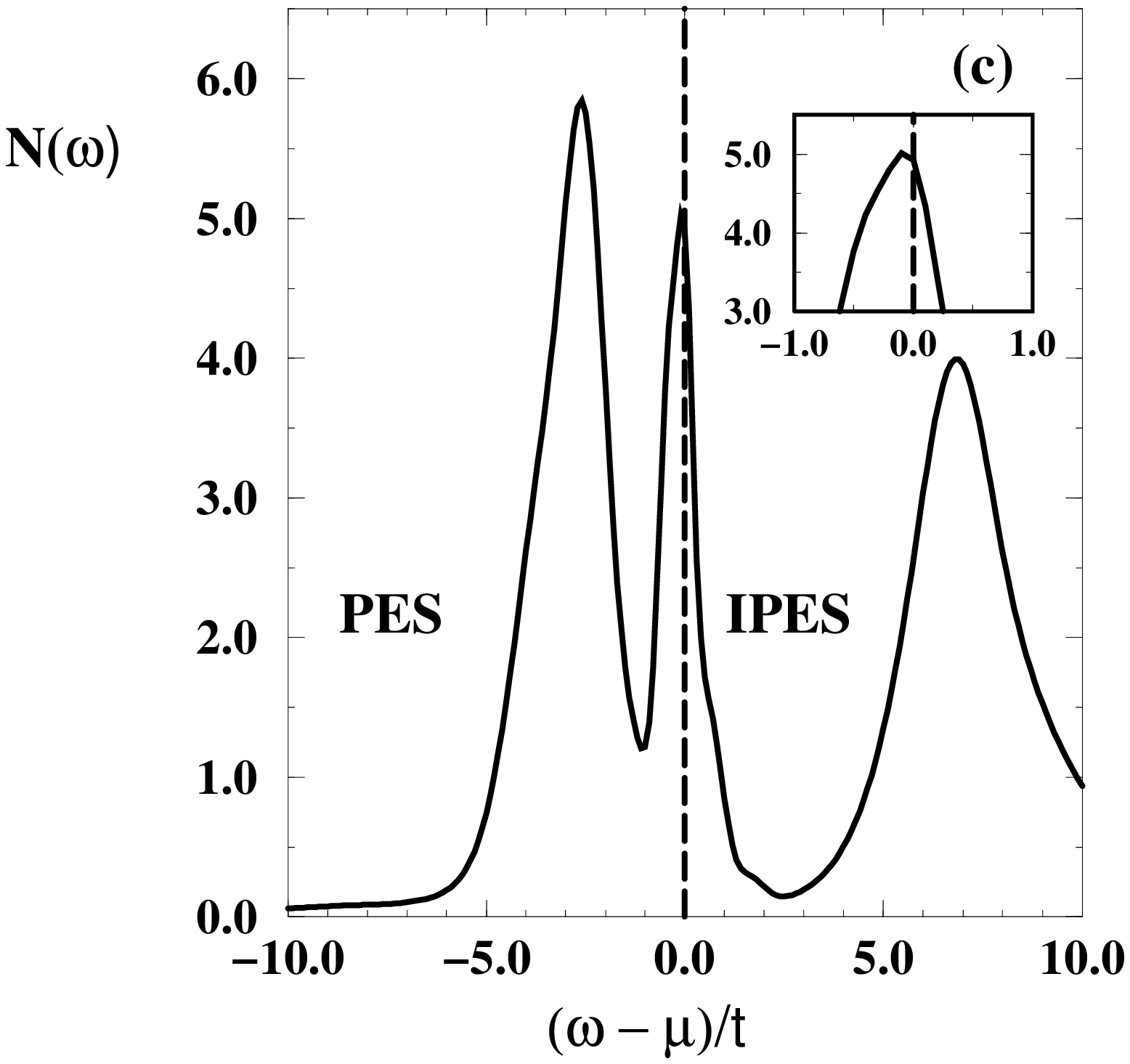,height=5.6cm,bbllx=132pt,bblly=155pt,bburx=570pt,bbury=578pt,angle=270}}
\caption{(a) Spectral functions $A({\bf p},\omega)$ of the $U-t-t'$ Hubbard
model at $U/t=10$, $t'/t=-0.35$, and $T=t/3$, using QMC-ME techniques on
a $6 \times 6$ cluster. The density is $\langle n \rangle = 0.94$. From
the bottom, the momenta are along the main diagonal from $(0,0)$ to
$(\pi,\pi)$, from there to $(\pi,0)$, and finally back to $(0,0)$; (b)
Energies of the dominant peaks in the spectral functions. The results in
the PES region are obtained from a two-gaussian analysis of the QMC-ME
results. The squares correspond to the incoherent part of the spectrum
(INC) and its error bars are not shown. The full circles form the
quasiparticle (qp) band, and their diameter is proportional to the
intensity. The error bars are given by the width at half the height of
the gaussian corresponding to the q.p.  peak at each momenta. The upper
Hubbard band is also shown (open circles) without error bars; (c)
density of states $N(\omega)$ obtained by summing the $A({\bf
p},\omega)$'s.}
\label{fig2}
\end{figure}

It is interesting to note that the q.p. band includes momentum
$(\pi,\pi)$, although with a very small 
intensity. This is reasonable for two reasons: (i) first, the
presence of a charge gap in the spectrum, as clearly seen in Fig.1b,
suggests that bands starting, say, at $(0,0)$ cannot simply disappear
at some other momentum by 
crossing $\mu$. By continuity they have to extend all along the 
Brillouin zone (unless their q.p. weight vanishes); (ii) in addition,
the presence of strong antiferromagnetic correlations implies a doubling of the unit
cell of dynamical origin and thus ``shadow band'' features should appear
in the spectrum, as discussed in several models with strong
antiferromagnetic correlations\cite{bob,shadow,shadow2,onellion2,preuss2,langer,2Dshadow,haas2}. 
In other words, AF introduces 
an extra symmetry in the problem that links ${\bf p}$ with ${\bf p} + (\pi,\pi)$. 
These shadow features are certainly
weak compared to the rest of the band, but are nevertheless present in the
spectrum of Fig.1b.

The rest of the spectral weight is contained in the broad incoherent
feature at energies deeper than the q.p. peak, and also in the UHB. The
latter presents some nontrivial momentum dependence near $(\pi,\pi)$.
This behavior seems a remnant of the results for noninteracting
electrons.  Finally, Fig.1c contains the density of states (DOS)
$N(\omega) = \sum_{\bf p} A({\bf p},\omega)$ at half-filling.  The gap
$\Delta \sim 5t$ and the sharp q.p. band can be clearly identified.

\subsection{Finite Hole Density}

%\subsection{\langle n \rangle = 0.94}

Fig.2 contains the QMC-ME results obtained at density
$\langle n \rangle =0.94$, with
the rest of the parameters as in Fig.1. In Fig.2a the spectral weight
$A({\bf p},\omega)$ is shown. There are several interesting features 
in this result. For example, the upper Hubbard band has lost weight
compared with results at $\langle n \rangle = 1.0$.
This is not surprising since as the hole density grows
the chances that an electron injected in the lattice will populate
already occupied sites is reduced.

The most interesting physical consequences of hole doping are obtained
in the PES region. Here the chemical potential has moved to the top of
the valence band, more specifically into the q.p. band.
Fig.2b shows the results of a two-peak analysis of the PES region
similar to that performed at half-filling. Interpolating results along
the main diagonal it is observed that the q.p. peak at
 momentum ${\bf p}=(\pi/2,\pi/2)$ 
should be
 approximately at the chemical potential for this density, while $(\pi,0)$ is 
still about $2t$ below. The q.p. band has distorted its shape
in such a way that considerable weight has been moved above $\mu$ in 
the vicinity of $(\pi,\pi)$. Actually, now the q.p. band resembles a more
standard tight-binding dispersion although with a renormalized
hopping amplitude  smaller than the bare one.
Note, however, that
the position of the q.p. peak at momentum $(0,0)$ is somewhat
pathological since it is located above the chemical potential. We do not
have an explanation for this anomalous behavior,

\begin{figure}[t]
\centerline{\psfig{figure=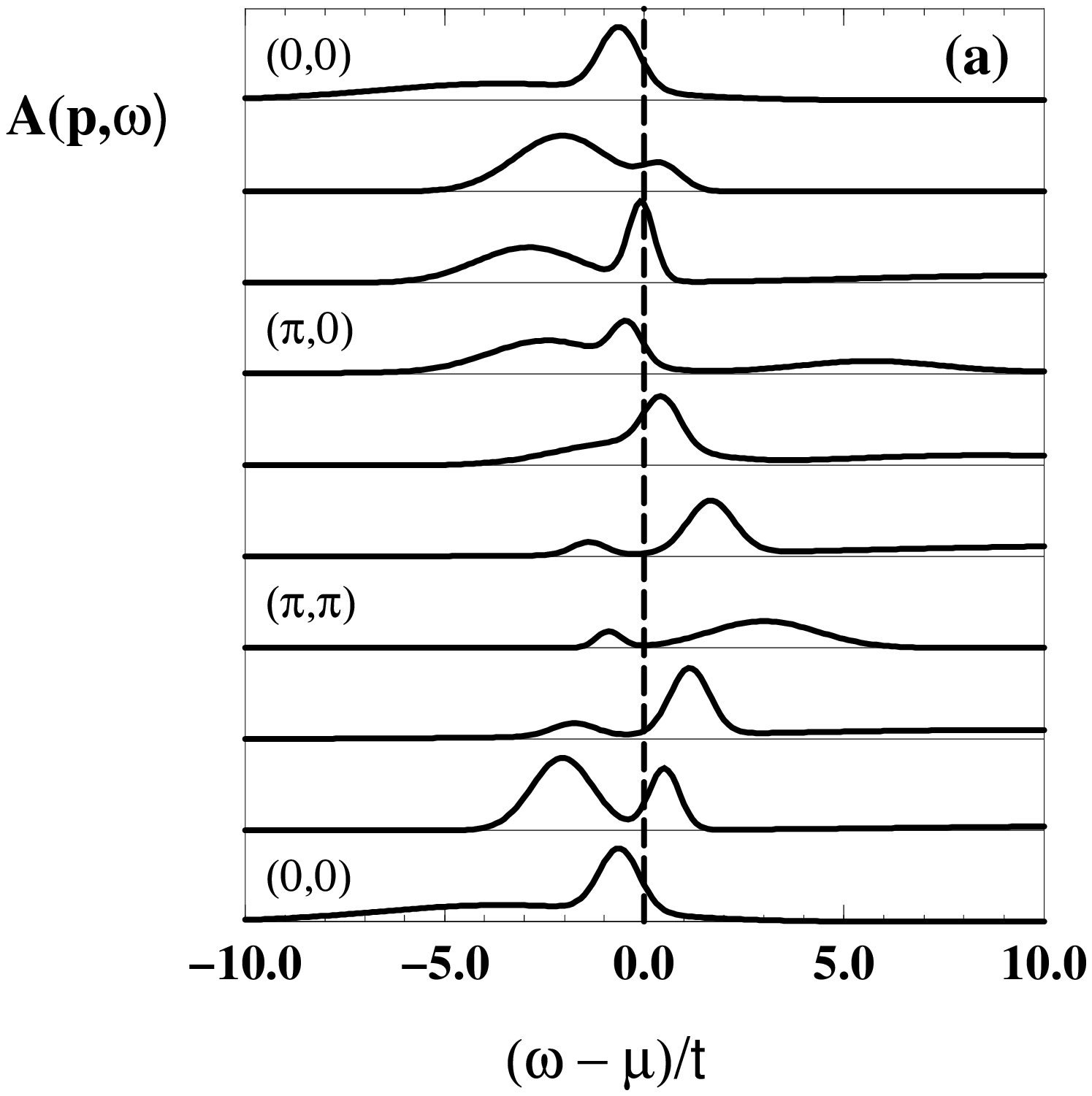,height=5.7cm,bbllx=132pt,bblly=155pt,bburx=570pt,bbury=578pt,angle=270}}
\centerline{\psfig{figure=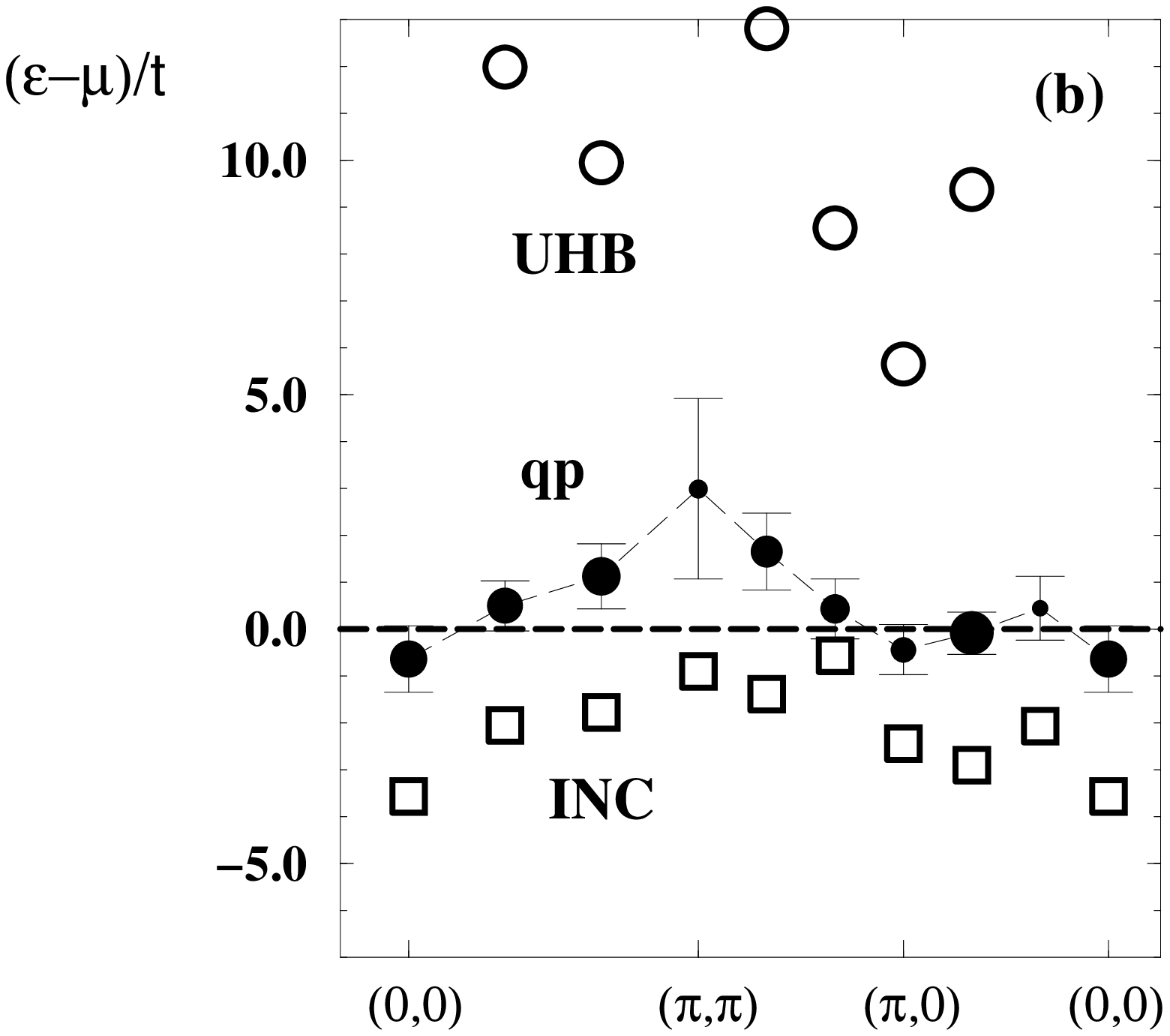,height=5.7cm,bbllx=132pt,bblly=155pt,bburx=570pt,bbury=578pt,angle=270}}
\centerline{\psfig{figure=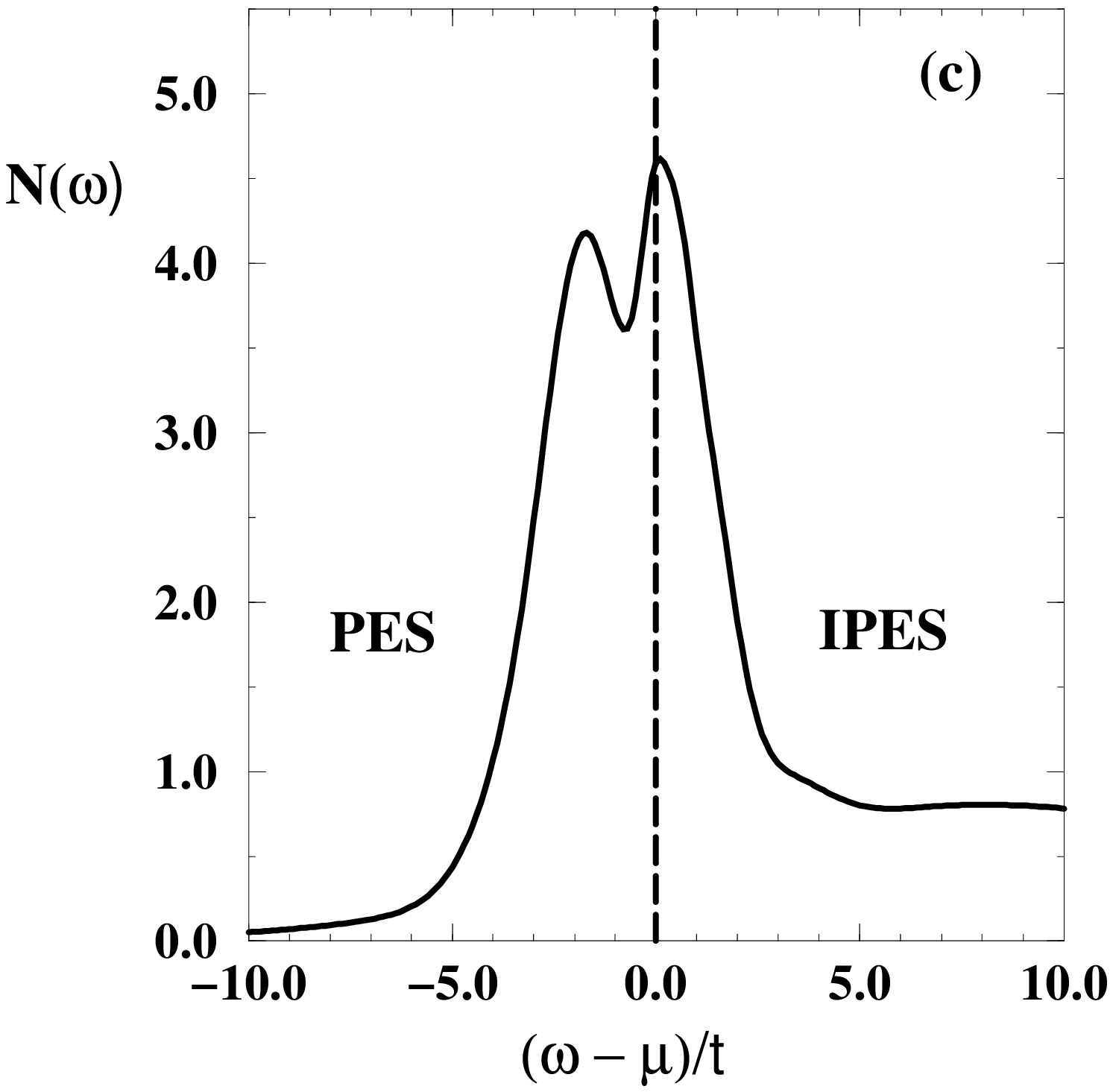,height=5.7cm,bbllx=132pt,bblly=155pt,bburx=570pt,bbury=578pt,angle=270}}
\caption{(a) Spectral functions $A({\bf p},\omega)$ of the $U-t-t'$ Hubbard
model at $U/t=10$, $t'/t=-0.35$, and $T=t/3$, using QMC-ME techniques on
a $6 \times 6$ cluster. The density is $\langle n \rangle = 0.84$. From
the bottom, the momenta are along the main diagonal from $(0,0)$ to
$(\pi,\pi)$, from there to $(\pi,0)$, and finally back to $(0,0)$; (b)
Energies of the dominant peaks in the spectral functions. The results in
the PES region are obtained from a two-gaussian analysis of the QMC-ME
results. The squares correspond to the incoherent part of the spectrum
(INC) and its error bars are not shown. The full circles form the
quasiparticle (qp) band, and their diameter is proportional to the
intensity. The error bars are given by the width at half the height of
the gaussian corresponding to the q.p.  peak at each momenta. The upper
Hubbard band is also shown (open circles) without error bars; (c)
density of states $N(\omega)$ obtained by summing the $A({\bf
p},\omega)$'s.}
\label{fig3}
\end{figure}

\begin{figure}[t]
\centerline{\psfig{figure=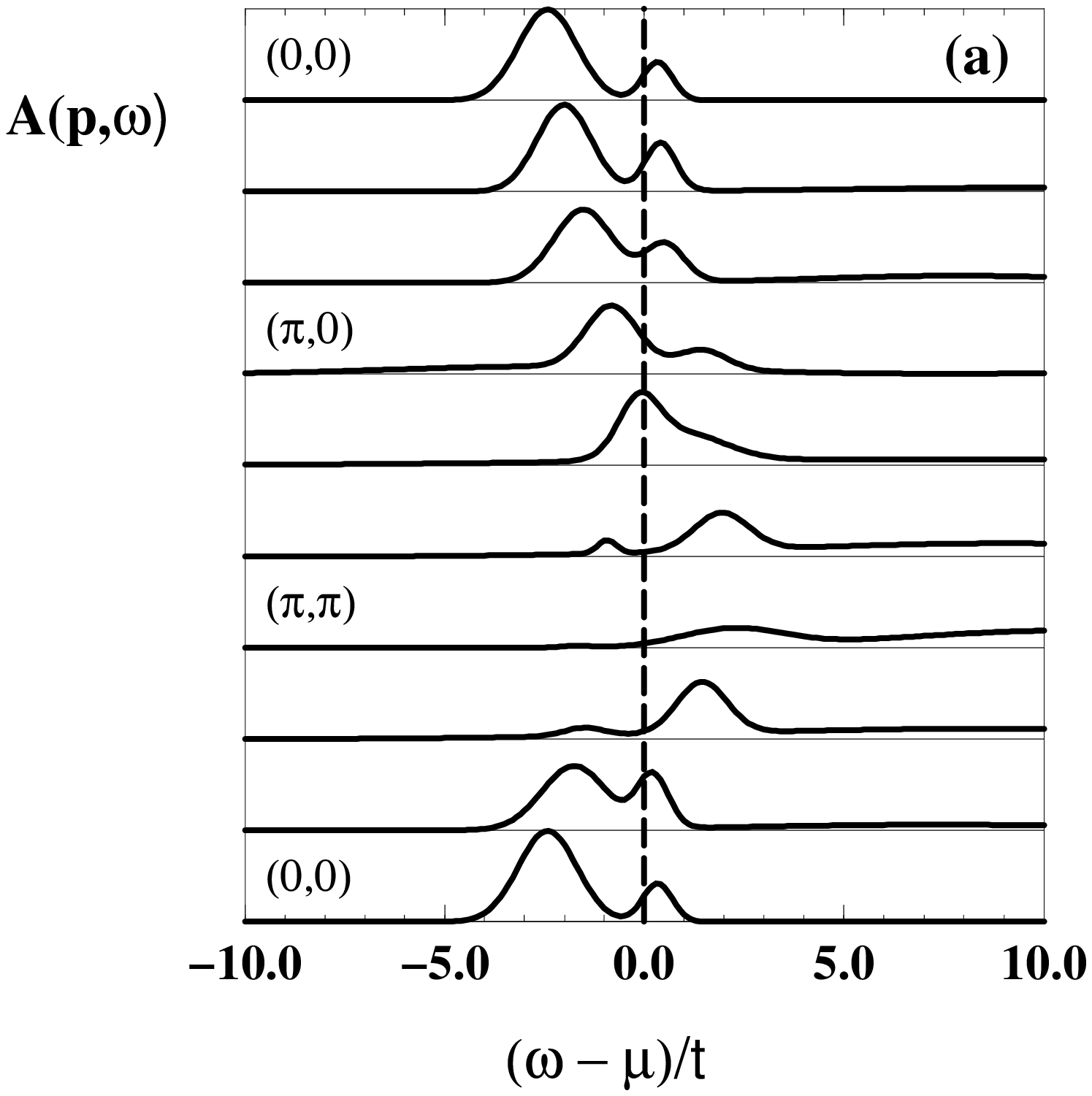,height=5.7cm,bbllx=132pt,bblly=155pt,bburx=570pt,bbury=578pt,angle=270}}
\centerline{\psfig{figure=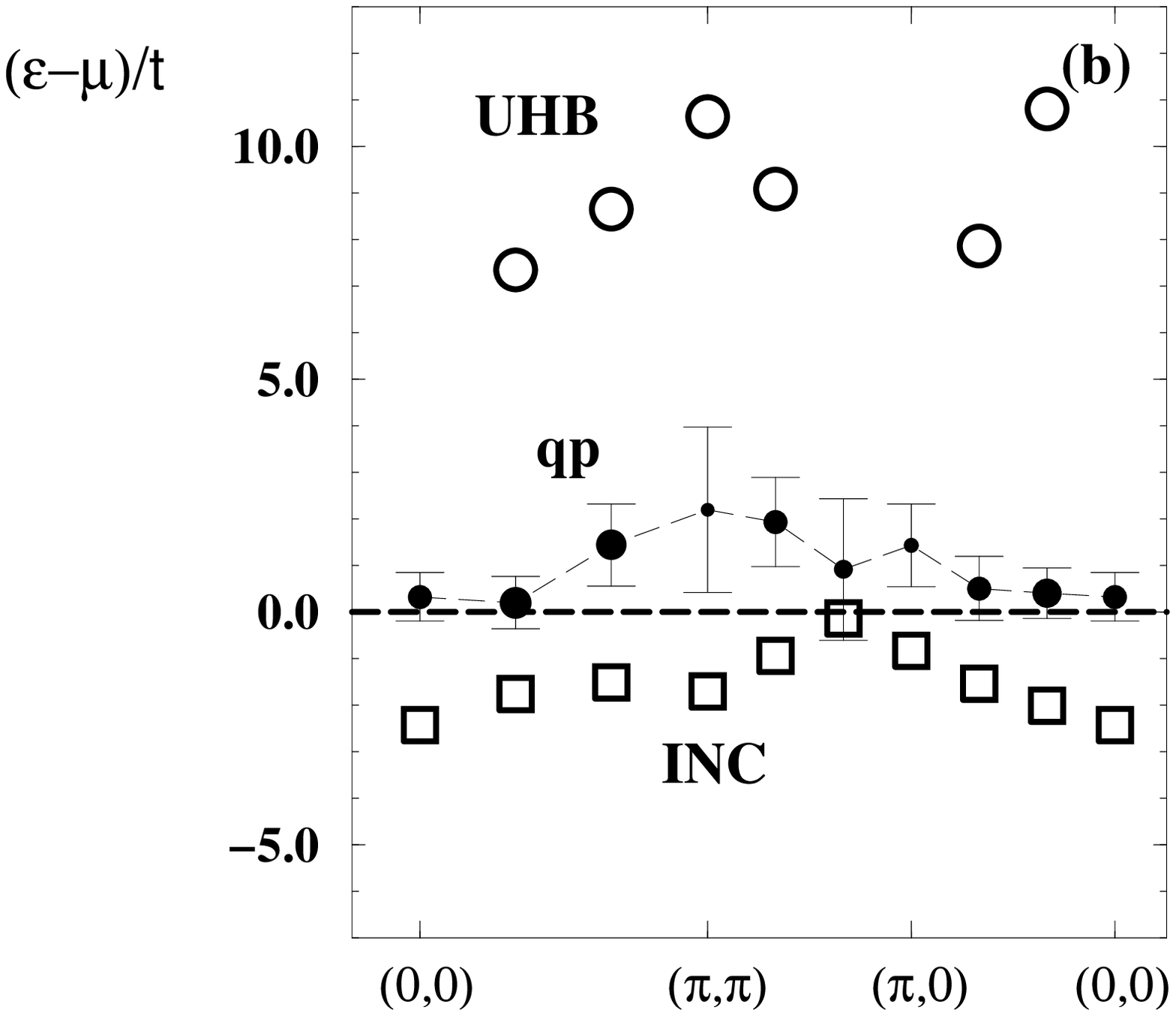,height=5.7cm,bbllx=132pt,bblly=155pt,bburx=570pt,bbury=578pt,angle=270}}
\centerline{\psfig{figure=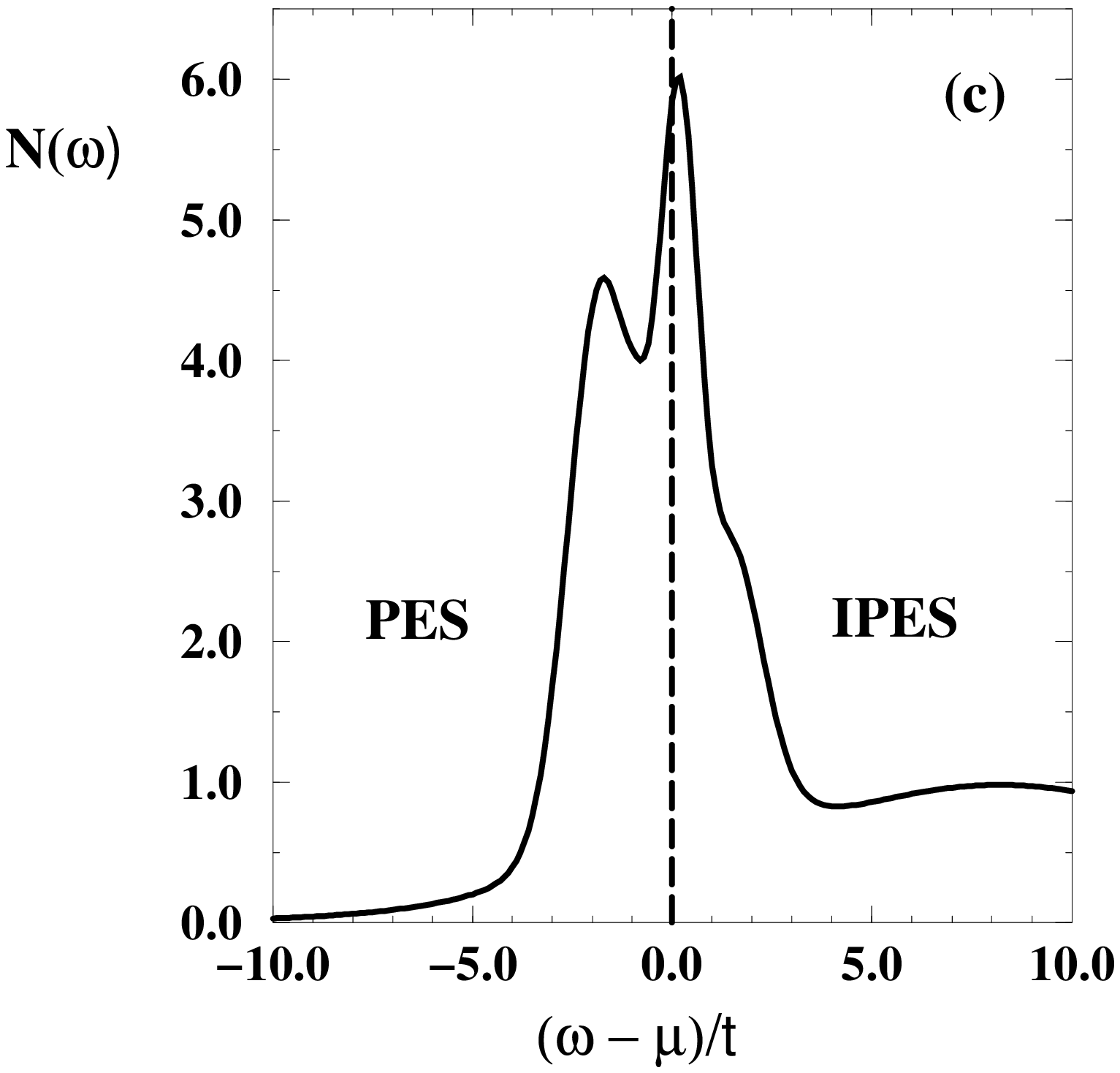,height=5.7cm,bbllx=132pt,bblly=155pt,bburx=570pt,bbury=578pt,angle=270}}
\caption{(a) Spectral functions $A({\bf p},\omega)$ of the $U-t-t'$ Hubbard
model at $U/t=10$, $t'/t=-0.35$, and $T=t/3$, using QMC-ME techniques on
a $6 \times 6$ cluster. The density is $\langle n \rangle = 0.78$. From
the bottom, the momenta are along the main diagonal from $(0,0)$ to
$(\pi,\pi)$, from there to $(\pi,0)$, and finally back to $(0,0)$; (b)
Energies of the dominant peaks in the spectral functions. The results in
the PES region are obtained from a two-gaussian analysis of the QMC-ME
results. The squares correspond to the incoherent part of the spectrum
(INC) and its error bars are not shown. The full circles form the
quasiparticle (qp) band, and their diameter is proportional to the
intensity. The error bars are given by the width at half the height of
the gaussian corresponding to the q.p.  peak at each momenta. The upper
Hubbard band is also shown (open circles) without error bars; (c)
density of states $N(\omega)$ obtained by summing the $A({\bf
p},\omega)$'s.}
\label{fig4}
\end{figure}

\begin{figure}[t]
\centerline{\psfig{figure=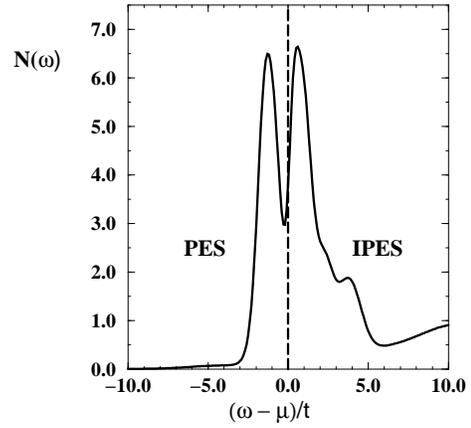,width=5.7cm,bbllx=132pt,bblly=155pt,bburx=570pt,bbury=578pt,angle=270}}
\caption{Density of states $N(\omega)$ corresponding to the $U-t-t'$
Hubbard model at $U/t=10$, $t'/t=-0.35$, and $T=t/3$, using QMC-ME
techniques on a $6 \times 6$ cluster. The density is $\langle n \rangle
= 0.65$.}
\label{fig5}
\end{figure}

\noindent which likely is caused
by the Maximum Entropy procedure. The
incoherent part of the spectrum (INC) is also clearly visible in the calculation. At
this density it remains entirely filled, i.e. the electrons removed from
the system have been taken from the q.p. band. Finally, note that the shadow
features are no longer prominent. This is correlated with a
substantial reduction of the antiferromagnetic correlation length $\xi_{AF}$
 at this density and temperature compared with the results at 
half-filling. This is not in contradiction with claims that
recently observed features in ARPES are induced by
antiferromagnetism\cite{shadow,shadow2} since
at very low temperatures $\xi_{AF}$ likely remains robust for a
larger hole density region near half-filling than it occurs at the
relatively high temperature $T=t/3$.
In Fig.2c the DOS for $\langle n \rangle = 0.94$ is shown. The q.p. band is  sharp, and $\mu$
lies slightly to the right of the top of this band 
(see inset of Fig.2c). A pseudogap 
generated by the large coupling $U/t$ is still clearly visible.

%\subsection{\langle n \rangle = 0.85}

Fig.3a contains the QMC-ME results obtained at $\langle n \rangle = 0.84$.
Compared with the results at $\langle n \rangle = 0.94$, now $\mu$
is located deeper inside the q.p. band according to the two-gaussians
analysis presented in Fig.3b. Interpolations clearly show that
  ${\bf p}=(\pi/2,\pi/2)$ 
for this band
is now in the IPES region, while $(\pi,0)$ is at or very close to
$\mu$ but still in the PES regime. The incoherent weight in PES remains robust.
Fig.3c shows the DOS at this density. $\mu$ is located slightly to the
left of the q.p. band maximum, the pseudogap has virtually melted, and there is
only a tiny trace of the upper Hubbard band.
The presence of $(\pi,0)$ very close to $\mu$ is correlated with
experimental results showing ``flat bands'' 
near the Fermi energy at optimal doping\cite{flat-exper,flat,langer,otherflat}.

Fig.4 contains the QMC results obtained at $\langle n \rangle =0.78$. This
density is particularly interesting since now $\mu$ seems to have
crossed most of  the q.p. band\cite{comm2}. This can be observed directly in
the spectral weight (Fig.4a), and in its two-peak analyzed output (Fig.4b).
This detail will have interesting consequences for some theories
of high-

\begin{figure}[t]
\centerline{\psfig{figure=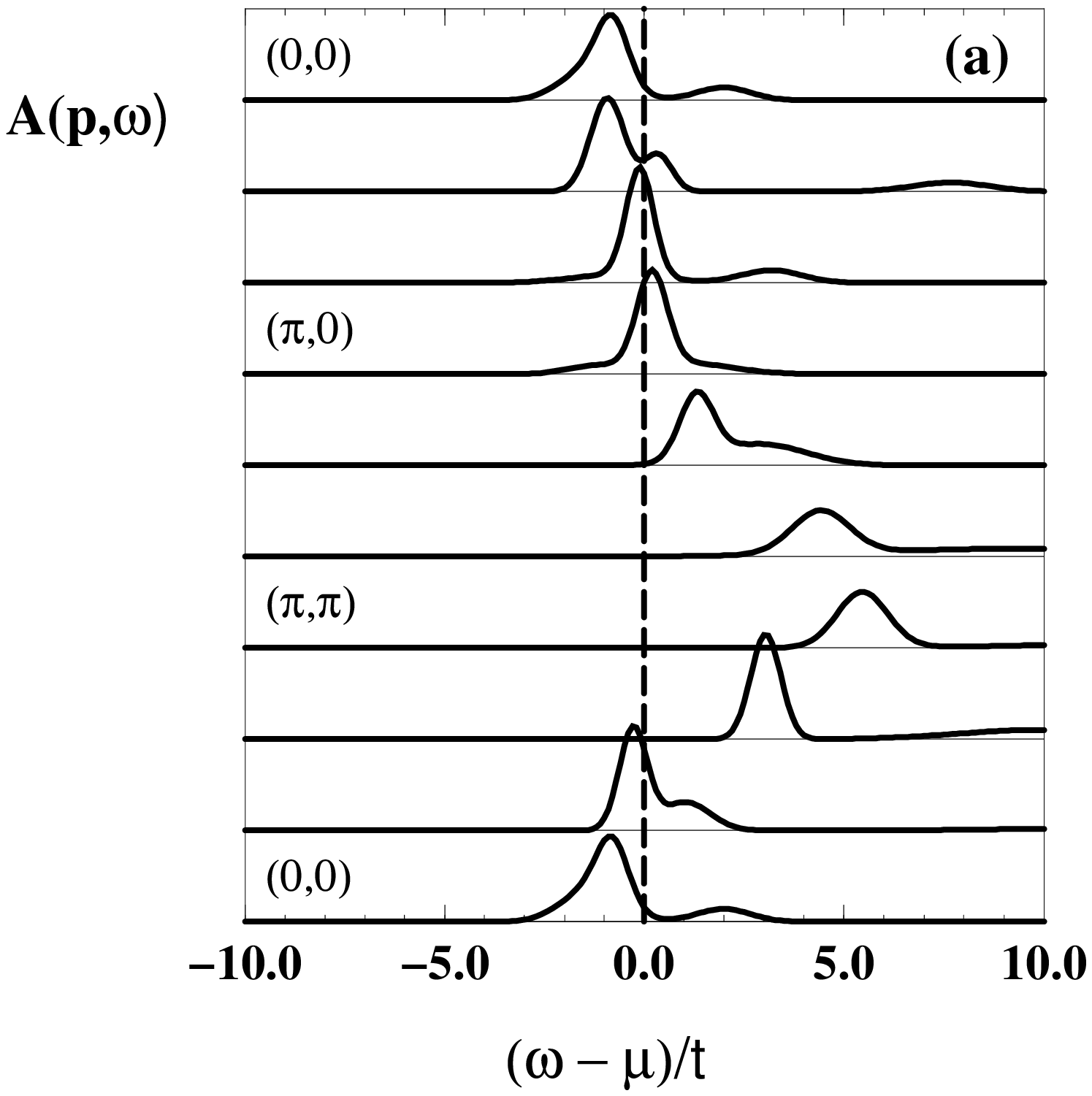,width=6.2cm,bbllx=132pt,bblly=155pt,bburx=570pt,bbury=578pt,angle=270}}
\centerline{\psfig{figure=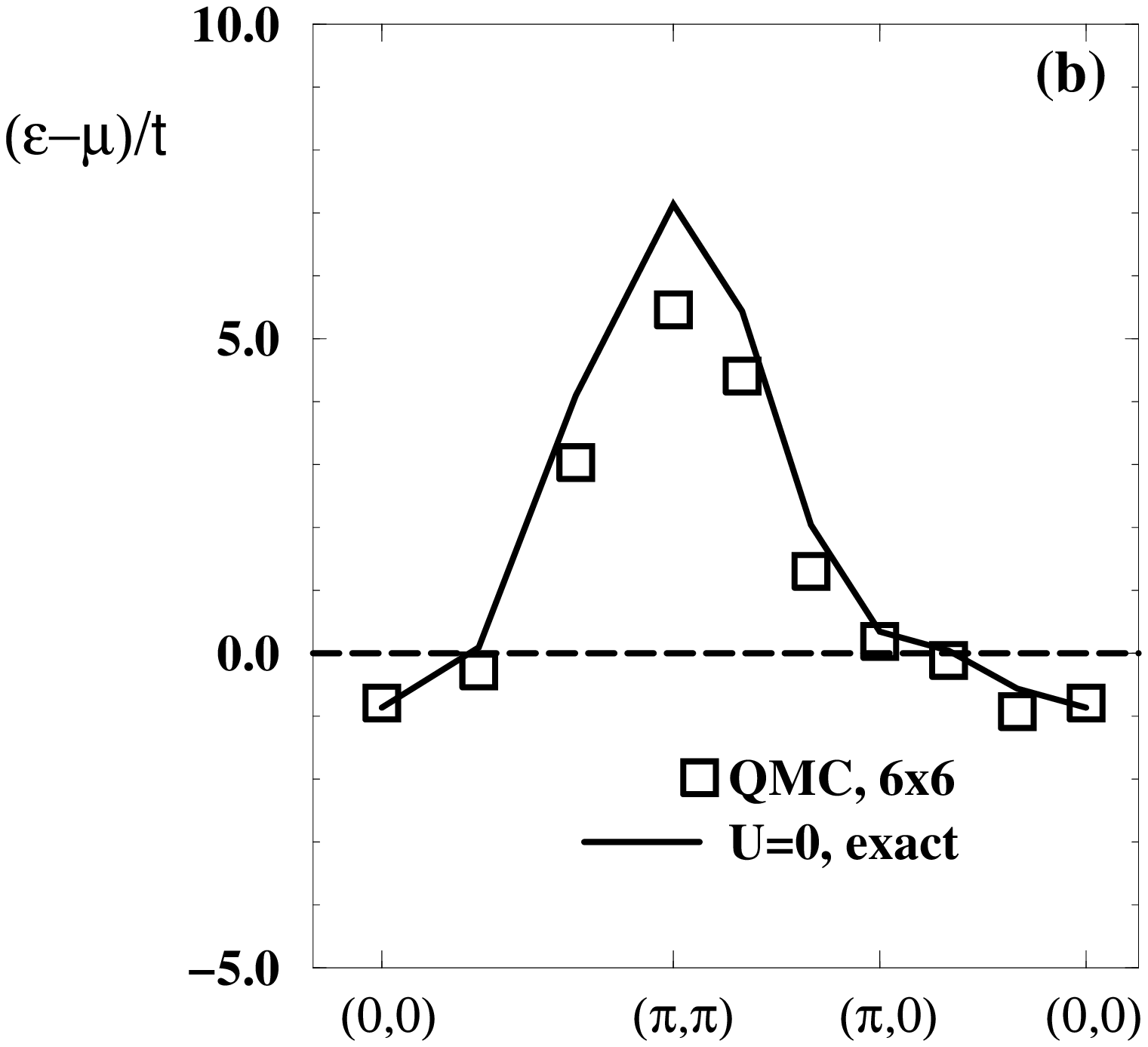,width=6.2cm,bbllx=132pt,bblly=155pt,bburx=570pt,bbury=578pt,angle=270}}
\centerline{\psfig{figure=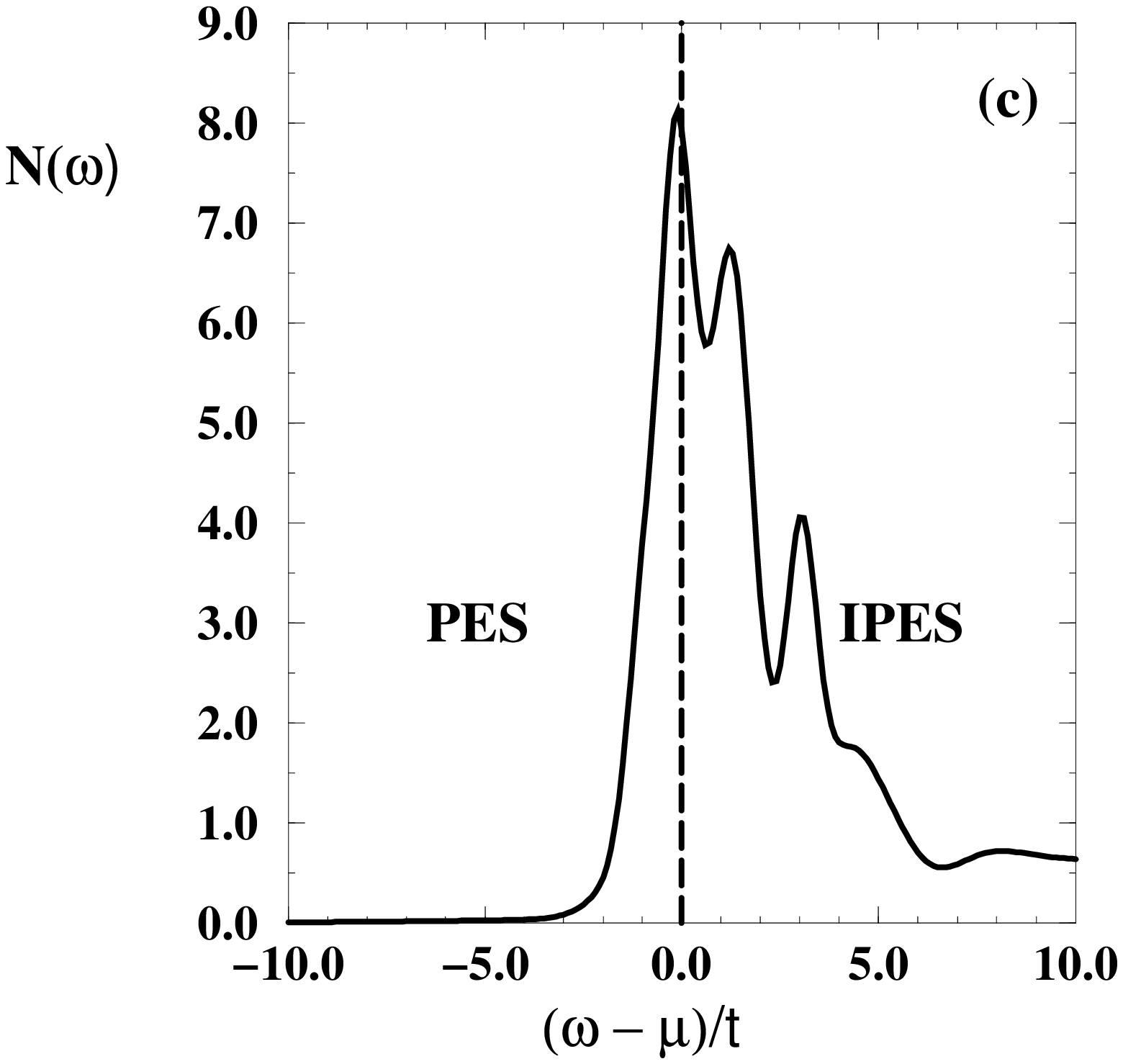,width=6.2cm,bbllx=132pt,bblly=155pt,bburx=570pt,bbury=578pt,angle=270}}
\caption{(a) Spectral functions $A({\bf p},\omega)$ of the $U-t-t'$ Hubbard
model at $U/t=10$, $t'/t=-0.35$, and $T=t/3$, using QMC-ME techniques on
a $6 \times 6$ cluster. The density is $\langle n \rangle = 0.47$. From
the bottom, the momenta are along the main diagonal from $(0,0)$ to
$(\pi,\pi)$, from there to $(\pi,0)$, and finally back to $(0,0)$; (b)
Energies of the dominant peaks in the spectral functions.  The solid
line is the dispersion in the noninteracting limit $U/t=0.0$; (c)
density of states $N(\omega)$ obtained by summing the $A({\bf
p},\omega)$'s.}
\label{fig6}
\end{figure}

\noindent $T_c$ uprates, as described later in this paper.
The DOS is shown in Fig.4c. The same trend is also
observed at $\langle n \rangle = 0.65$ for which here only the
DOS is shown (Fig.5). At this density the q.p. band is empty 
 and the INC becomes sharper presumably
due to its proximity to the chemical potential. It is interesting to
note that the chemical potential
in Fig.5 is located at a sharp minimum in the density of states.
%separating features that evolved with density from the 
%q.p. band and incoherent part of the spectrum at half-filling.

Results at $\langle n \rangle = 0.47$ (Fig.6) show that
the q.p. band has melted 
and at this low density the spectral weight is dominated by 
features that can be traced back to the incoherent part of the PES
spectrum at half-filling (Fig.6a).
In Fig.6b the position of the dominant
peak is given as a function of momentum, and in Fig.6c the DOS is shown. It is interesting that now
the results are nicely fitted by the non-interacting tight-binding
dispersion (solid line in Fig.6b).
Thus, the ``free electrons'' limit is approximately recovered 
in the strong coupling regime at ``quarter-filling''. Again, note that
the peak structure in the spectrum 
seems to have emerged from the incoherent
part of the valence band observed
at half-filling.

\section{Comparison of QMC-ME results with ARPES experiments}

As discussed in the Introduction, ARPES results have shown that the
quasiparticle peak at ${\bf p} = (\pi,0)$ is very sensitive to hole
doping at least in the underdoped regime\cite{marshall,ding}.  For the
same densities, the q.p. dispersion along the main diagonal from $(0,0)$
to $(\pi,\pi)$ does not change as much. In this section it will be
investigated if the numerical results of Sec.III are compatible with the
ARPES data.

In Fig.7, we schematically show what occurs
upon doping when a tight binding dispersion
including  a NN hopping amplitude $t^*$, and a
NNN amplitude $t'^*$ with a ratio 

\begin{figure}[h]
\centerline{\psfig{figure=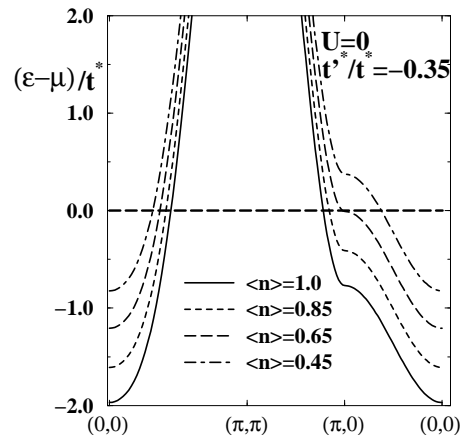,width=6.1cm,bbllx=132pt,bblly=155pt,bburx=570pt,bbury=578pt,angle=270}}
\caption{Density evolution of a non-interacting band, referred to the
chemical potential, as its density changes (shown in the figure). For
details see the text.}
\label{fig7}
\end{figure}

\begin{figure}[h]
\centerline{\psfig{figure=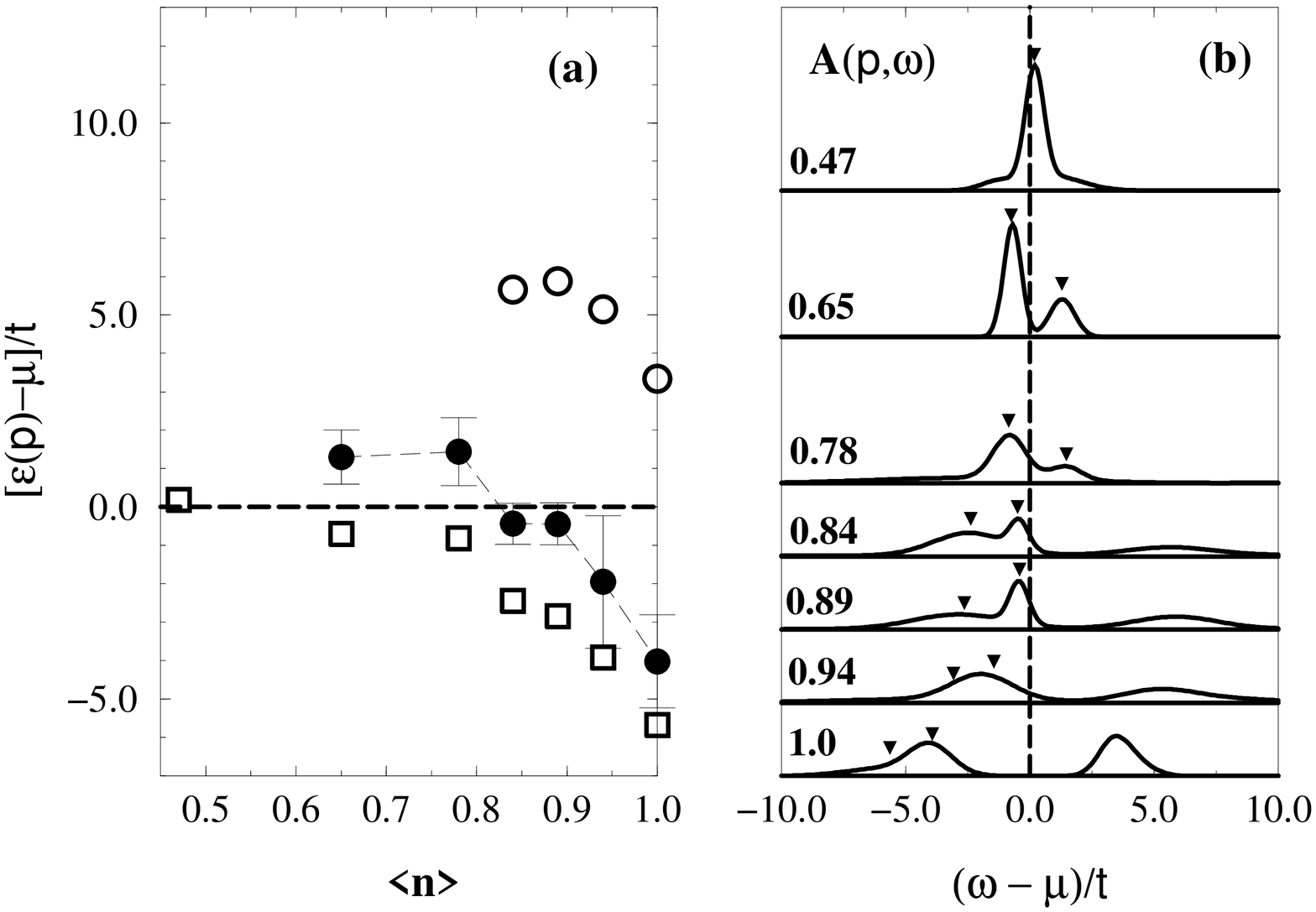,width=5.0cm,bbllx=90pt,bblly=200pt,bburx=565pt,bbury=650pt,angle=270}}
\caption{(a) Energy of the quasiparticle (referred to the chemical
potential) in units of $t$ for the $U-t-t'$ Hubbard model using the
parameters of Figs.1-6.  The results here correspond to momentum ${\bf
p} = (\pi,0)$. The solid circles, squares and open circles are the q.p.,
INC and UBH positions for this momentum; (b) quasiparticle spectral
function corresponding to momentum ${\bf p}=(\pi,0)$ for several
densities $\langle n \rangle$ (indicated). The small triangles show the
positions of the dominant peaks in the two-gaussian analysis of the PES
data.}
\label{fig8}
\end{figure}

\begin{figure}[b]
\centerline{\psfig{figure=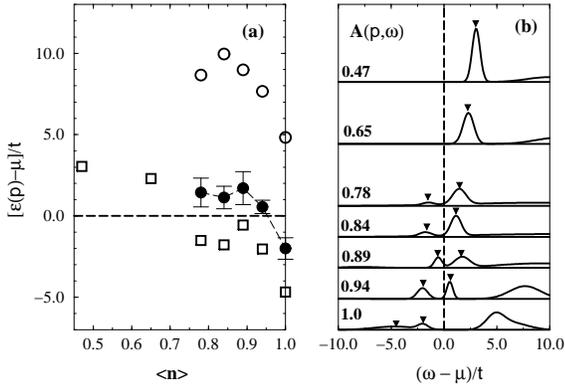,width=5.0cm,bbllx=140pt,bblly=200pt,bburx=565pt,bbury=650pt,angle=270}}
\caption{Energy of the quasiparticle (referred to the chemical
potential) in units of $t$ for the $U-t-t'$ Hubbard model using the
parameters of Figs.1-6.  The results here correspond to momentum ${\bf
p} = (2\pi/3,2\pi/3)$. The solid circles, squares and open circles are
the q.p., INC and UBH positions for this momentum; (b) quasiparticle
spectral function corresponding to momentum ${\bf p} =(2\pi/3,2\pi/3)$
for several densities $\langle n \rangle$ (indicated). The small
triangles show the positions of the dominant peaks in the two-gaussian
analysis of the PES data.}
\label{fig9}
\end{figure}

\noindent $t'^*/t^* = -0.35$ 
is used. The reason for this exercise is that
the QMC-ME results of Sec.III have shown that
rapidly upon hole doping the quasiparticle dispersion
resembles that of a renormalized non-interacting set of
electrons. Then, it is interesting to analyze what would
occur with such dispersion as the density {\it within the
q.p. band} changes (i.e. the density $\langle n \rangle$
shown in Fig.7 must be considered as the filling of the
q.p. band, rather than the density of the whole system).
It is clear from the figure
that $(\pi,0)$ is intrinsically more sensitive to doping than the main
diagonal. Actually the results resemble in part those of Marshall et
al.\cite{marshall} specially the experimental 
ARPES dispersion corresponding to
slightly underdoped 

\begin{figure}[t]
\centerline{\psfig{figure=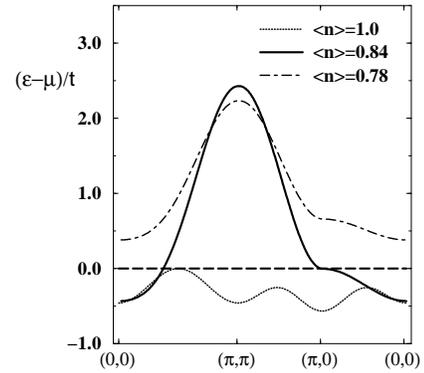,width=5.3cm,bbllx=140pt,bblly=155pt,bburx=570pt,bbury=578pt,angle=270}}
\caption{Best fits of the QMC-ME quasiparticle dispersion corresponding to
densities $\langle n \rangle = 0.84$ and $\langle n \rangle = 0.78$.
The results at $\langle n \rangle = 1.0$ are taken from
Ref.[10]arbitrarily locating the top of the band at the
chemical potential.}
\label{fig10}
\end{figure}

\noindent samples of Bi2212 with $T_c = 85K$, as well as previous
experiments for optimally doped samples\cite{flat-exper}, where the
q.p. band crosses the chemical potential between $(\pi,\pi)$ and
$(\pi,0)$.  This lead us to conjecture that the ARPES experiments in the
slightly underdoped regime may have observed the hole filling of a
``free-like'', but narrow, q.p. band.  If this speculation is correct,
then it is here predicted that $(\pi,0)$ will eventually cross the
chemical potential when the system becomes overdoped, as Fig.7 shows. To
the best of our knowledge there are no ARPES results in this
regime. Experimental work for overdoped cuprates would clarify the issue
of whether the ``flat bands'' found in the optimal regime remain locked
near $\mu$ or smoothly cross the Fermi energy as the hole doping
increases. Our results favor the latter, at least within the resolution
of the QMC-ME methods.

Further evidence that $(\pi,0)$ travels across the chemical potential is
given in Fig.8a,b. There  $A({\bf p},\omega)$ with ${\bf p}=(\pi,0)$, 
taken from
Figs.1a-6a, is shown again adding also results for
densities $\langle n \rangle = 0.89$ and $0.65$
for completeness. The peak at this momentum
belonging to the q.p. band crosses
$\mu$ at a density $\langle n \rangle \approx 0.82$. 
Then, as the hole density grows
a movement up in energy of the
${\bf p} = (\pi,0)$ quasiparticle is clearly observed, as in
 ARPES experiments\cite{sawa}.
At a larger
hole density, such as quarter-filling, the q.p. peak, now in the IPES regime,
substantially reduces 
its weight and disappears. 
At this density the feature associated with the incoherent part of the spectrum at
half-filling is now located close to the Fermi energy, contributing to
the dispersion that resembles the $U=0$ result (Fig.6b). 
The UHB rapidly looses
weight moving away from half-filling.
Fig.9a,b show similar results but for ${\bf p} = (2\pi/3,2\pi/3)$, which 
is representative of the behavior along the main diagonal on the $6 \times
6$ cluster. The q.p. associated to this momentum
crosses the chemical potential at a smaller hole density ($\langle n \rangle
\approx 0.95$).

Summarizing the results of this section, 
in Fig.10 three representative q.p. dispersions obtained with QMC-ME
methods (Sec.III)
are shown. For the half-filled case the results
are accurately known from the self-consistent Born 
approximation\cite{nazarenko}, and $(\pi,0)$ is clearly below the
chemical potential, which
here is arbitrarily located at the top of the band. 
Note that the intensity of the q.p. peak is much weaker near $(\pi,\pi)$
than for other momenta at half-filling.
For small hole
doping densities, such as $\langle n \rangle = 0.84$, a ``free-like'' but
narrow
quasiparticle band is observed in QMC simulations (also in studies of
the $U-t$ model\cite{bulut2,moreo})
 and $(\pi,0)$ is now at the Fermi energy. 
%Note that the region
%from $(0,0)$ to $(\pi/2,\pi/2)$ has not changed as much, 
%in agreement with experiments.
The vicinity of $(\pi,\pi)$ is the region
affected the most by hole doping
due to the reduction of antiferromagnetic
correlations. Finally, at lower densities, such 
as $\langle n \rangle = 0.78$, the chemical potential has crossed
the q.p. band and now $(\pi,0)$ is above the Fermi energy.
% a dispersion similar to the
%$U=0$ result is obtained. Here, the quasiparticle peak at
%$(0,\pi)$ that evolves from half-filling has negligible intensity
%and it is well above $\mu$. However, note that the ... has generated
%a new sharp peak at this momentum that lies close to the Fermi energy.
%Both peaks should not be confused with each other.
We consider
that the scenario depicted in Fig.10 for the density evolution of the
q.p. dispersion can provide a simple qualitative explanation for the
ARPES results found experimentally. It predicts that the q.p. peak at
$(\pi,0)$ will eventually disappear from the ARPES signal into the IPES region, an
effect that could be tested experimentally. At densities intermediate
between $\langle n \rangle = 1.0$ and $0.84$, the peak
at $(\pi,0)$ should evolve smoothly between these two limits, and thus
close to half-filling ``pseudogap'' features will likely appear in
simulations if lower temperatures could be reached. Reducing the
temperature below $T=t/3$ is particularly
important for the generation of strong spin fluctuations which are
crucial to induce the quasiparticle dispersion at half-filling shown in
Fig.10. Note that ``shadow'' features should actually be present at $\langle n
\rangle = 0.84$ but they are too weak to be detected by QMC-ME methods,
and probably also by ARPES experiments which have large backgrounds
in their signals. Thus, the existence of hole pockets cannot be shown
from the current Monte Carlo simulations available.
%However, certainly additional
%work and more
%accurate many-body techniques are needed to fully confirm our ideas.

%Alternative explanations for the behavior of $(0,\pi)$
%based on the presence of preformed pairs in the normal state have 
%been discussed\cite{marshall}. 
%These ideas
%are reasonable and we are currently investigating them. 
%To distinguish
%d-wave preformed pairs from the results of Fig.11  extra
%efforts should be devoted to the study of point $P$ in the figure
%located along the $(0,\pi)$ to $(\pi,\pi)$ line where the low hole density
%q.p. peak crosses
%$\mu$. If this crossing is actually detected then the present ideas would
%become a serious candidate to explain the ARPES data. However, if no
%crossing is experimentally found, implying that the q.p. starting
%at $(0,\pi)$ somehow bends down in energy when moving towards the direction
%of $(\pi,\pi)$, preformed pairs approaches would be favored. We encourage 
%experimentalists to addresses this issue, as well as the crossing of the
%Fermi energy by the flat bands as predicted above to clarify if
%preformed pairs ideas are necessarily needed to explain the 
%high-Tc normal state properties.

\section{Implications of QMC results
for some  theories of High-$T_c$}

In this section the implications of our QMC results for recently
proposed scenarios for high-$T_c$ cuprates will be discussed.
Two features found
in the simulation will be important, namely the ``tight-binding'' shape
of the q.p. band at finite hole density (although with renormalized
parameters) and the crossing of the q.p. band by $\mu$ for
densities in the approximate
range $0.70 \leq \langle n \rangle \leq 1.0$.

\subsection{Large peak in the DOS}

As explained in the introduction, the presence of ``flat'' regions
in the experimental normal-state q.p. dispersion is a remarkable feature of the
phenomenology of hole-doped cuprates\cite{flat-exper}. These
flat bands  are located around momenta
${\bf p} = (\pi,0)$ and $(0,\pi)$, and at optimal doping they are
${\rm \sim 10meV}$ below the Fermi energy\cite{flat-exper}. 
Studies of holes in the 2D $t-J$ and Hubbard models at and away 
from half-filling have suggested that antiferromagnetic correlations may play
an important role in the generation of these
features\cite{flat,langer,otherflat,bulut2}. Our results for the $U-t-t'$ model
suggest the presence of flat bands in the regime close to $\langle n \rangle =
0.84$, although care must be taken with the 
 finite resolution of the ME-generated q.p. peaks.
But even if these flat regions were not quantitatively described 
by one band electronic models, the 
intrinsic small bandwidth of the q.p. band at small hole density
could be enough to induce a large peak in the DOS which can be used to 
enhance the superconducting
critical temperature, once a source of hole attraction is identified. 
This leads to a natural explanation for the existence of an
``optimal doping''  which in this framework 
occurs when the peak in the q.p. band is reached by $\mu$\cite{afvh}.
It is also natural to label as ``underdoped'' the regime where $\mu$
is to the right of the ``flat band'' peak (i.e.
at higher energies), and ``overdoped'' when it is to
the left (i.e. at lower energies)\cite{foot3}. 

\begin{figure}[t]
\centerline{\psfig{figure=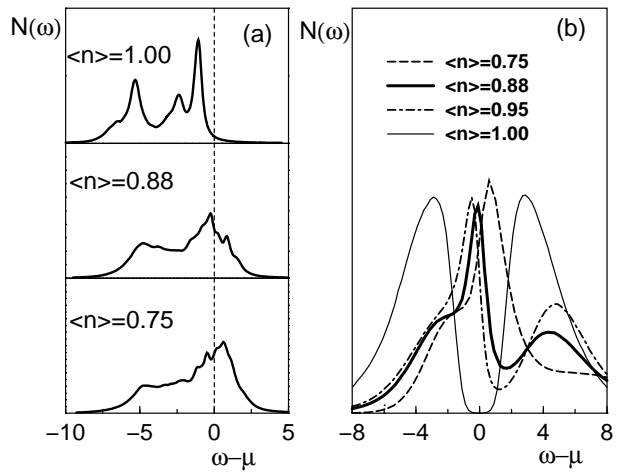,width=4.5cm,bbllx=20pt,bblly=200pt,bburx=620pt,bbury=620pt,angle=270}}
\caption{(a) $N(\omega)$ for the 2D $t-J$ model obtained with exact
diagonalization techniques averaging results for clusters with 16 and 18
sites, at $J/t = 0.4$ and for the densities indicated.  The
$\delta$-functions were given a width $\eta = 0.25t$. Similar results
were found at other values of $J/t$; (b) $N(\omega)$ for the one band
Hubbard model obtained on a $4 \times 4$ cluster using QMC and ME, but
without reducing the coefficient of the entropy as in Sec.III. The
temperature is $T=t/4$ and $U/t=12$.  Densities are indicated.}
\label{fig11}
\end{figure}

Previous analytical calculations\cite{afvh}
{\it assumed} the survival of the large DOS obtained at half-filling as
the density of holes grows.
The results of our simulations (Sec.III)
allow us to study the evolution of the DOS with hole density
and judge if these ideas are realistic.
The presence of a robust
q.p. peak is certainly confirmed by our results, and Figs.1b-4b show
that as $\langle n \rangle$ decreases from 1, the q.p. band is crossed
by $\mu$ as conjectured before\cite{afvh}.
The DOS of the standard $t-J$ and Hubbard models
(with $t'=0$) also have a large peak in the DOS, as found in
previous numerical simulations. For completeness, here those
results are also presented.
In Fig.11a,
$N(\omega)$ for the 2D $t-J$ model obtained with exact
diagonalization (ED) techniques is shown at several densities\cite{review}.
At half-filling, a large DOS peak appears at the top of the valence 
band\cite{flat}. 
Note that substantial
weight exists at energies
far from $\mu$, i.e. the large peak carries only a fraction
of the total weight, in agreement with the QMC simulations of Sec.III. 
In the $t-J$ model, the maximum
in the DOS is not strictly located at the top of 
the valence band but at slightly
smaller energies\cite{flat}. This effect is enhanced by
adding $t'$ hopping terms to the $t-J$ Hamiltonian, as shown 
in the simulation results
of Sec.III. As
$\langle n \rangle$ decreases, the peak in Fig.11a is now much 
broader but it remains well-defined. 
At $\langle n \rangle \sim 0.88$, $\mu$ 
is located close to the energy where
$N(\omega)$ is maximized. At $\langle n \rangle \sim 0.75$, $\mu$ moves to the left
of the peak. These results are quantitatively similar
to those of our QMC simulations for the $U-t-t'$ model, and also with
results for the $t'/t=0.0$ Hubbard model (Fig.11b). 
Thus, while accurate extrapolations to
the bulk limit are difficult, the simple qualitative picture emerging from
these studies, namely that strong correlations generate a q.p. peak
in the DOS which is crossed by $\mu$, seems robust.

As explained before, this q.p.-peak-crossing of the DOS is important
since once a source of hole attraction exists in the system,
a superconducting (SC) gap would open at $\mu$, and  
the resulting $T_c$ could be enhanced due to the large number of states
available. The numerical results, both QMC and ED, 
are thus compatible with scenarios where a large $T_c$ is
obtained due to an increase in $N(\mu)$. 
%In the
%context of the Hubbard and $t-J$ models the sharp q.p.
%peak itself is induced by AF correlations, i.e. the effect
%is caused by strong correlations rather than by the use of
%peculiar band structures.
Since the peak width 
increases substantially with hole density, strictly speaking
the rigid band filling of the half-filled hole dispersion
is invalid. However, such an approximation seemed to
have captured part of the qualitative physics of the problem,
 since the DOS peak is not washed out by
a finite hole density.

%Note that this possible SC gap would not alter
%much the already robust peak in the DOS,
%even with $T_c \sim 100K$.

In the cases discussed here, i.e. the $U-t-t'$ and $t-J$ models,
 calculations of the 
spin-spin correlations show that 
$\xi_{AF}$  is approximately a couple of lattice
spacings when $\mu$ is located near the DOS peak, 
becoming smaller as the overdoped regime is reached. 
Thus, a nonzero $\xi_{AF}$
and $\mu$ near a large DOS peak are $correlated$ features.
It is in this respect that scenarios where AF correlations produce
a large DOS that enhances $T_c$\cite{afvh} differ from
vH theories where divergences in the DOS are caused by band
effects already present before interactions are switch on\cite{vh}.

While the existence of a robust peak in the DOS is in good agreement
with ARPES data\cite{flat-exper}, it is in apparent disagreement with 
specific heat 
studies for YBCO\cite{loram}.
The lack of ${\bf p}$-resolution in the
specific heat measurements may solve this puzzle. Actually,
angle-$integrated$ PES results for Bi2212 do not show the sharp flat
features found in ARPES for the same material\cite{fujimori,imer}.
Similar effects may affect the specific heat data which should be
reanalyzed to search for DOS large peaks.

\subsection{Kondo resonances vs AF induced quasiparticle}

Previous QMC-ME studies of the Hubbard model for
$t'/t=0.0$\cite{bulut2,bulut1,shadow2} reported results
qualitatively
similar to those shown in Fig.11b, where the DOS  obtained
on a $4 \times 4$ cluster is presented for the one band Hubbard model
at $U/t=12$, and  $T=t/4$. The ME technique used in Fig.11b 
is the same as in those previous
simulations, and it does not have the resolution of the present
ME variation discussed in Sec.III. 
The crossing of a peak in the DOS by $\mu$ is clearly
observed in these ME
simulations but
at half-filling there are  no q.p. peaks and
Fig.11b can thus be naively interpreted as the  
``generation'' by doping of a Kondo-like
peak at the top of the valence band
which does not exist at $\langle
n \rangle =1$.
%, as it occurs in $D=\infty$ calculations.
However, our current results (Fig.1c) 
%using 
%ME methods which have higher resolution
actually show that at $\langle n \rangle = 1.0$ a well-defined
q.p. peak is present, as predicted by a variety of studies of
the zero temperature $t-J$ model with one hole\cite{review,moreo}. Also
experimentally in the cuprates it has
been already established\cite{fujimori}  that the states observed  in PES
upon doping are already present in the insulator and are $not$ Kondo
resonances. Thus, Figs.1b-4b provide evidence that the q.p. features
observed at $\langle n \rangle < 1$ are smoothly connected 
to peaks already present at half-filling.
Similar conclusions have also been obtained
in geometries other than the 2D square lattice, such as a $t-J$
ladder\cite{haas2,troyer}.

\subsection{$d_{x^2 - y^2}$ in the Hubbard model}

The results of Figs.1b-6b show that the q.p. band is very
sensitive to doping, at least at the temperatures and couplings
used in this QMC-ME study. In particular a few percent hole doping
is enough to transform the half-filling dispersion, containing the
extra symmetry induced by long-range AF order, into a
tight-binding-like dispersion although with a small hopping amplitude.
These results are similar to others previously published in the
literature. For example in Fig.12a,b the 
q.p. dispersion  at $\langle n \rangle =1$, and at
$\langle n \rangle  \sim 0.87$ for the $t-J$
model is reproduced from Ref.\cite{moreo} (for Hubbard model results,
see Ref.\cite{bulut2,ortolani}). Upon doping, vestiges of the flat regions 
remain, inducing a large peak in the DOS (see Fig.11).
However, the region around ${\bf p} = (\pi, \pi)$ has changed
substantially, i.e. the AF shadow region observed at
$\langle n \rangle =1$
reduced its intensity and 
considerable weight was transferred to the IPES region.
The q.p. dispersion at $\langle n \rangle \sim 0.87$ of the $t-J$ model
can also be fitted by a tight-binding 
nearest-neighbors (NN) dispersion with a small effective hopping
likely associated with $J$. Then, both previous literature results and
the present simulations agree on the qualitative aspects of the 
evolution with doping of $A({\bf p},\omega)$.

The changes in the q.p. dispersion with hole doping can be interpreted
in two ways: First, note that $A({\bf p}, \omega)$ is influenced by
matrix elements of $bare$ fermionic operators connecting states with $N$
and $N\pm 1$ particles. This is

\begin{figure}[h]
\centerline{\psfig{figure=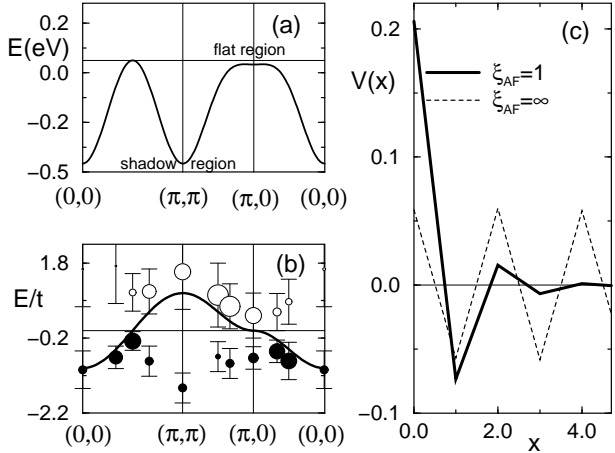,width=4.5cm,bbllx=20pt,bblly=200pt,bburx=620pt,bbury=620pt,angle=270}}
\caption{(a) q.p. energy vs momentum obtained at half-filling using the
$t-t'-J$ (from Refs.[12,32]).  The result shown, that we call
$\epsilon_{AF} ({\bf p})$, is a good fit of Monte Carlo data on a $12
\times 12$ cluster at $J/t = 0.4$; (b) q.p. dispersion vs momentum at
$\langle n \rangle = 0.87$ and $J/t = 0.4$ using exact diagonalization
of 16 and 18 sites clusters for the $t-J$ model (from
Ref.[25]). The open (full) circles are IPES (PES) results.
Their size is proportional to the peak intensity. The solid line is the
fit $\epsilon_{NN}({\bf p})$ described in the text; (c) $V(x)$ along the
x-axis after Fourier transforming the smeared potential $V({\bf p}) =
\delta({\bf p} - {\bf Q})$ (see text). $\xi_{AF}$ is given in lattice
units.}
\label{fig12}
\end{figure}

\noindent important when the q.p. weight is small,
i.e. when the state $c_{{\bf p} \sigma} |gs \rangle_N$ does not have a
large overlap with
the ground state of the $N-1$ particles subspace $|gs \rangle_{N - 1}$ 
($|gs \rangle_N$
being the ground state with $N$ particles). 
If the 
hole excitation is instead created by a  new operator $\gamma_{{\bf p}\sigma}$
that incorporates the dressing of the hole by spin fluctuations,
then $\gamma_{{\bf p}\sigma} |gs \rangle_N$ may now have
a large overlap with $|gs \rangle_{N-1}$\cite{qp,eder}.
In other words, if the dressed hole state
resembles an extended spin polaron, then the physics deduced from PES
studies, which rely on the sudden removal of a bare electron from the
system, may be misleading. To the extent that spin polarons remain
well-defined at finite density, the use of $\gamma_{{\bf
p}\sigma}$ will induce spectral weight rearrangements
between the PES and IPES regions, and 
the results at small hole density could resemble those at
half-filling after such spectral weight redistribution
takes place.
If this idea were correct, then the study of superconductivity
and transport, both regulated by $dressed$ quasiparticles, could
indeed be handled by filling 
a rigid band given by $\epsilon_{AF}({\bf p})$.
It is likely that this idea will work
in the underdoped regime where $\xi_{AF}$ is robust, and results for
2D clusters\cite{eder}  and ladders\cite{jose3} already
support these claims.
Note that the fact that the
bandwidth at finite hole density remains much smaller than $8t$ shows
that strong correlations still play an important role in the dispersion.

However, an alternative is that the difference between Figs.12a and 12b
for the $t-J$ model and Figs.1b and 2b for the $U-t-t'$ model correspond
to an intrinsic change

\begin{figure}[h]
\centerline{\psfig{figure=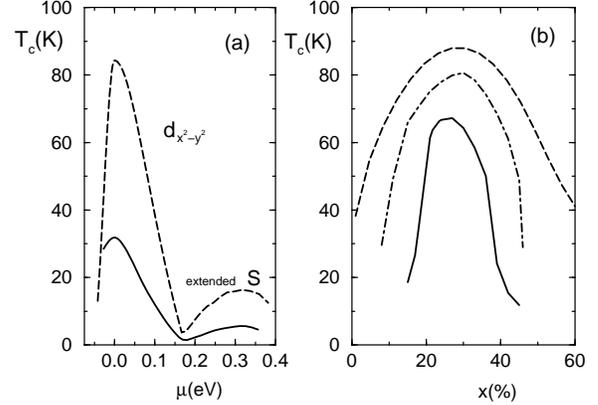,width=4.5cm,bbllx=20pt,bblly=200pt,bburx=620pt,bbury=620pt,angle=270}}
\caption{(a) $T_c$ vs $\mu$ obtained with the BCS gap equation, with
$\mu=0$ as the chemical potential corresponding to the saddle-point.
The dashed line corresponds to results using $\epsilon_{AF}({\bf p})$,
and the solid line to $\epsilon_{NN}({\bf p})$. Interpolations between
these two extreme cases can be easily constructed using two band models
with weights regulated by $Z$-factors (actually we calculated $T_c$
using a two-pole Green's function in the gap equation and the result
smoothly interpolated between those shown in the figure).  Note the
presence of both ${ d_{x^2 - y^2}}$-wave and extended s-wave SC; (b)
$T_c$ for d-wave SC vs the percentage $x$ of filling of the q.p. band
(i.e. $not$ of the full system).  The dashed line is the same as in
(a). The solid line corresponds to the BCS gap equation result making
zero the weight $Z_{\bf p}$ of states in the q.p. dispersion that are at
energies from the saddle point larger than 2.5\% of the total bandwidth
(i.e. basically including states only in a window of energy $\sim 125K$
around the flat regions). The dot-dashed line is the same but using a
window of $\sim 250K$ around the flat regions.}
\label{fig13}
\end{figure}

\noindent in the q.p. dispersion as $\xi_{AF}$
decreases. This is the less favorable case for the real-space pairing
approaches\cite{afvh} which are 
constructed at half-filling,
and thus we should analyze this
possibility in detail here. For this purpose,
we have applied the standard BCS formalism
to an effective
 model with a low density of quasiparticles having a dispersion
$\epsilon_{NN}({\bf p})/eV = -0.2(cosp_x + cosp_y)$ (assuming
$t=0.4eV$), which roughly
reproduces the dominant features of Fig.12b (the reader should not
confuse this dispersion with that of free particles whose bandwidth is four
times larger. A large amount of incoherent weight is still observable at
the density of Fig.12b although it is not shown explicitly. The same
occurs in Fig.2b for the $U-t-t'$ model). As described
before, to study $T_c$ we should
include a NN attraction induced by AF between these 
quasiparticles. While naively it may seem
dubious to use the same interaction both at and away from
half-filling, the hole-hole potential should not be much
affected at
distances shorter than $\xi_{AF}$. This can be illustrated by a
real space analysis (Fig.12c) of a smeared $\delta$-function potential
of AF origin $V({\bf q}) = 
\xi_{AF}/ [ 1 + \xi_{AF}^2 ( {\bf q} - {\bf Q} )^2 ]$,
where the lattice spacing is set to 1, and ${\bf Q} = (\pi,\pi)$.
Fig.12c shows that the NN potential ($x=1$)
does not change noticeably as $\xi_{AF}$ is reduced, while $V(x>1)$
is rapidly suppressed. 
Then, using the same
NN form of the potential for many densities
should not be a bad approximation\cite{comm7}. 

%Previous results within the framework of the BCS gap equation
%using a dispersion calculated at half-filling, as well as a NN hole
%attraction, produced a $T_c$ of order 100K and superconductivity
%in the $d_{x^2 - y^2}$ channel, in good agreement with experiments\cite{afvh}.
%However, results such as those obtained in our simulations Fig....
%as well as in previous studies of the $t-J$ and one band Hubbard
%model, suggest that the q.p. dispersion changes substantially with
%doping rapidly
%becoming a  renormalized tight-binding band away from half-filling.
%How will the use of such a dispersion affect the results obtained in
%previous analytical studies? 

To analyze the stability of previous calculations
 let us then use the
tight-binding dispersion $\epsilon_{NN}({\bf p})$
obtained from the numerical analysis in the BCS gap equation.
%For simplicity here we use a relation $\epsilon ....$ obtained
%from previous results for the $t-J$ model away from half-filling,
%but similar conclusions can be obtained using the $U-t-t'$ QMC
%outcomes. 
Solving numerically the gap equation,  $T_c$ is shown
in Fig.13a. As reported before\cite{afvh}, 
superconductivity at
$T_c \sim 80-100K$ in the ${ d_{x^2 - y^2}}$ channel 
appears naturally if the hole dispersion $\epsilon_{AF}({\bf p})$
 at half-filling is used. 
Nevertheless, if instead $\epsilon_{NN}({\bf p})$
is used, 
the vestiges of the  flat bands present in this narrow dispersion produce
$T_c \sim 30K$ which is still large\cite{grabo}. Even more remarkable is the fact 
that the ${ d_{x^2 - y^2}}$ character of the SC state 
is maintained. This result can be understood noticing that a combination of
$\epsilon_{NN} ( {\bf p})$ with
an attractive NN potential effectively locates the Hamiltonian in the 
family of ``$t-U-V$'' models with $U$ repulsive and $V$ attractive, where
it is known that for a ``half-filled'' electronic band the dominant SC
state is ${ d_{x^2 - y^2}}$-wave\cite{micnas}. 
In other words, when $\mu$ is at the flat region
of the $\epsilon_{NN}({\bf p})$ dispersion it approximately
corresponds to an effective ``half-filled'' q.p. band resembling a free electron
dispersion (but with smaller bandwidth), leading
to a ${ d_{x^2 - y^2} }$-wave SC state (Fig.13a).
Then, even if the q.p. dispersion changes substantially with
doping near the ${\bf Q}=(\pi,\pi)$ point, such an effect
does $not$ seem to alter the main qualitative features found in previous
studies\cite{afvh}. This is a remarkable result and it is caused by the
rapid depopulation of the q.p. band as the hole density grows. Then,
while $\langle n \rangle$ governs the actual global density of the system,
there is a subtle {\it hidden} density (i.e. the population of the q.p. band)
that may influence considerably on the physics of the problem. At $\langle
n \rangle = 0.78$ and for the $U-t-t'$ model, 
this q.p. band is nearly empty while globally the system
is still close to half-filling. While the use of $\epsilon_{AF} ( {\bf
p})$ or $\epsilon_{NN} ( {\bf p})$ is certainly a rough approximation, the
existence of a robust $T_c$ and d-wave superconductivity in both cases
suggests that similar results would be obtained using more realistic
models for the hole pairing interaction.

\subsection{Influence of shadow bands on SC}

The present analysis also shows that the
AF ``shadow'' regions of the half-filling hole dispersion
  $\epsilon_{AF}({\bf p})$
are not crucial for the success of the real-space approach.
Using $\epsilon_{NN}({\bf p})$, which
does not contain weight in PES near $(\pi,\pi)$, 
$T_c$ is still robust and the d-wave
state remains stable, as shown in the previous subsection. 
To establish this result more clearly, we analyzed $T_c$
using $\epsilon_{AF}({\bf p})$ 
but modulating the contribution
of each momentum with a ${\bf p}$-dependent weight $Z_{\bf p}$ in the
one particle Green's function.
We considered the special case where $Z_{\bf p}$ is
zero away from a window of total width $W$ centered at the
saddle point, which is located in the flat
bands region. Inside the window $W$, the weight is
maximum i.e. $Z_{\bf p} =1$.
Such a calculation also addresses indirectly possible concerns
associated with widths $\sim (\epsilon_{AF}({\bf p}) -\epsilon_F)^2$
that q.p. peaks would acquire away from half-filling in standard Fermi
liquids ($\epsilon_F$ is the Fermi energy). Results are shown in Fig.13b, for d-wave SC.
Note that even in the case where $W$ is as small as just 5\% of the
total bandwidth (itself already small of order $2J$), 
$T_c$ remains robust and close to $70K$. Then, it
is clear that the dominant contribution to $T_c$ comes from the flat regions
and the shape of the q.p. dispersion away from them has a secondary
importance for the success of the real-space approach. However, it is 
important to remark that the calculation described in this section
showing that shadow features are not much important for 
the actual value of $T_c$ does not mean
that AF correlations can be neglected. On the contrary, the whole 
Antiferromagnetic Van Hove scenario\cite{afvh} and other similar
approaches are based on the notion that pairing is caused by 
spin fluctuations\cite{chubu}.

\subsection{SC in the overdoped regime}

Finally, novel predictions obtained 
in the regime where the q.p. band is almost fully
 crossed by $\mu$ (as
observed in Sec.III for $\langle n \rangle$ between 0.78 and 0.66)
are here discussed. From the point of view of the q.p. band this regime
is ``dilute'' but, again, this should 
not be confused with the bottom of the whole spectrum
since a large amount of weight remains in the incoherent 
 part of $A({\bf p},\omega)$.
% I.e. from the point of view of
%the quasiparticle band the system is diluted at a large value of
%$\langle n \rangle$.
%%
%
%$\epsilon_{AF}({\bf p})$ 
%and $\epsilon_{NN}({\bf p})$, when $\mu$ reaches the
%bottom of these bands (which, again, should not be confused with 
%the bottom of the full hole spectrum since a
%large amount of incoherent weight lies at energies lower than those of 
%the q.p. band). 
At this density a standard
BCS gap equation analysis
applied to a q.p. dispersion either constructed at half-filling,
as in Ref.\cite{afvh}, or phenomenologically obtained from our
data ($\epsilon_{NN}({\bf p})$), and supplemented by nearest-neighbor attraction induced by
antiferromagnetism,
shows that extended $s$-wave SC dominates over 
${ d_{x^2 - y^2}}$-wave SC. This corresponds to the
``overdoped'' regime (Fig.13a) i.e. to an overall
density in the vicinity of $\langle n \rangle \sim 0.70$\cite{oldprl}. 
This change in the symmetry of the SC state
can be understood recalling once again that the tight-binding dispersion
of the q.p., even including renormalized amplitudes,
supplemented by a NN attraction,
formally corresponds to an effective ``$t-U-V$'' model. It is 
well-known that in this model the
SC state symmetry changes from d- to s-wave as the
density is reduced away from half-filling to a nearly empty
system\cite{micnas,dago5}.
Actually the bound state of two particles on an
otherwise empty lattice with a NN tight binding
dispersion and NN attraction is s-wave.
To the extend that the AF- or NN-dispersions survive up to 
$\sim 25\%$ hole doping, 
as suggested by the numerical data of Sec.III as well as
previous literature(Ref.\cite{moreo} and references therein),
scenarios based on the real-space interaction of q.p.'s
predict a competition between 
extended s-wave and ${ d_{x^2 - y^2}}$-wave 
SC in the $overdoped$ regime.
Recent calculations
based on the analysis of the low electronic density $\langle n
\rangle \ll 1$ limit of the $t-J$ model led to analogous
conclusions\cite{kolte}. 
Our approach is based on
a very different formalism but it arrives to similar results,
and, thus, a crossover from d-
to s-wave dominated superconductivity in overdoped cuprates could occur.
Indeed recent ARPES data for overdoped Bi2212 
have been interpreted as corresponding to a mixing of s- and d-wave
states\cite{ma}. More work should be devoted to this potential
competition between d- and s-wave for overdoped cuprates.

\section{Summary}

In this paper the results of
an extensive numerical study of the $U-t-t'$ one band
Hubbard model with $t'/t=-0.35$ have been presented. The regime of strong
coupling $U/t=10$ was analyzed. These parameters are fixed
to reproduce ARPES data for the AF insulator. With maximum
entropy techniques the spectral function $A({\bf p},\omega)$
was studied for several electronic densities. It was observed
that  as the hole density grows away from
half-filling, the quasiparticle
band acquires ``non-interacting'' features although with
bandwidths substantially smaller than for the $U=0$ limit.
As the hole density increases the q.p. peak at $(\pi,0)$
rapidly changes its position relatively to the Fermi energy,
in qualitative agreement with 
recent ARPES experiments\cite{marshall,ding}. In the overdoped
regime it is predicted that the flat bands should be crossed
by the chemical potential and thus they should no longer be
observed in ARPES studies. The present results
also have implications
for some theories of high-$T_c$. The narrow
q.p. band produces a large peak in
the DOS induced by strong correlations, which
survives the presence
of hole doping and it can be used to boost $T_c$ once a source
of hole attraction is found. It is remarkable that both
with the hole dispersion found at half-filling or the one
observed using QMC-ME at, e.g., $\langle n \rangle = 0.84$,
superconductivity in the $d_{x^2 - y^2}$ channel dominates.
This result gives support to  
real-space pairing theories of
high-$T_c$\cite{afvh}, showing that the main ideas of the
scenario are stable upon the introduction of a finite hole
density.
The large DOS peak is crossed
by the chemical potential as $\langle n \rangle$ is reduced.
In the overdoped regime, i.e. when the q.p. band is almost
empty, a possible competition between extended-s and
$d_{x^2 - y^2}$ SC was discussed. 

\section{acknowledgments}

We thank A. Sandvik, M. Onellion, D. Dessau, J. C. Campuzano, and
Z. X. Shen for useful discussions.
D. D. is supported by grant ONR-N00014-94-1-1031.
A. M. and E.D. are supported by grant NSF-DMR-9520776. Additional 
support by the National High
Magnetic Field Lab and Martech is acknowledged.


\begin{references}
\bibitem{footnote} Note that the sign problem is most severe near optimal
doping.  Furthermore, the sign problem increases with coupling
strength and with the magnitude of $t'$.
\bibitem{foot3} Note that implicitly we are here
assuming that the $N(\omega)$ obtained 
by adding the numerically calculated $A({\bf p},\omega)$'s
can be related with the density of states that would appear in a BCS gap
equation. The difference resides in the matrix elements that modulate
the intensity in photoemission calculations.
\bibitem{comm7} Note also that the local character of the real-space 
potential does $not$ imply small Cooper pairs. Their size
can be regulated using both its range $and$ intensity.
\end{references}
\end{document}